\documentclass[namedreferences]{SolarPhysics} 
\usepackage[optionalrh]{spr-sola-addons} 
\usepackage{epsfig} 
\usepackage{graphicx}                    
\usepackage{amssymb}                    
\usepackage{color} 
\usepackage{url} 

\begin{document}

\begin{article}

\begin{opening}

\title{Automated Coronal Hole Detection using Local Intensity Thresholding Techniques}

\author{Larisza D.~\surname{Krista}$^{1}$\sep 
Peter T.~\surname{Gallagher}$^{1}$\ 
}

\runningauthor{Krista and Gallagher} 
\runningtitle{Automated coronal hole detection using EUV and X-ray intensity thresholding techniques}

\institute{$^{1}$Astrophysics Research Group, School of Physics, Trinity College Dublin, Dublin 2, Ireland.\\ 
email: \url{kristal@tcd.ie}\\ 
} 
\date{Received ; accepted }

\begin{abstract} {We identify coronal holes using a histogram-based intensity thresholding technique and compare their properties to fast solar wind streams at three different points in the heliosphere. The thresholding technique was tested on EUV and X-ray images obtained using instruments onboard STEREO, SOHO and {\it Hinode}. The full-disk images were transformed into Lambert equal-area projection maps and partitioned into a series of overlapping sub-images from which local histograms were extracted. The histograms were used to determine the threshold for the low intensity regions, which were then classified as coronal holes or filaments using magnetograms from the SOHO/MDI. For all three instruments, the local thresholding algorithm was found to successfully determine coronal hole boundaries in a consistent manner. Coronal hole properties extracted using the segmentation algorithm were then compared with {\it {\it in-situ}} measurements of the solar wind at $\sim$1~AU from ACE and STEREO. Our results indicate that flux tubes rooted in coronal holes expand super-radially within 1AU and that larger (smaller) coronal holes result in longer (shorter) duration high-speed solar wind streams.}

\end{abstract} 
\end{opening} 

\section{Introduction} 
\label{S-Introduction}

Understanding the connection between solar activity and space weather is important for accurately predicting geomagnetic disturbances that can affect radio communications, satellite and spacecraft operations, ground-based power systems and astronauts. Coronal mass ejections (CMEs) and flares are sources of energetic particles which can impact the Earth, and hence are of primary importance to space weather forecasting. As sources of high speed solar wind, coronal holes (CHs) are an important factor in producing recurring magnetic disturbances at Earth on time scales of days to months. It is well known that due to the longevity of CHs, they can drive important physical processes on a larger time-scale than transient events like CMEs. Over its lifetime, a coronal hole can input as much energy into the heliosphere as a CME \cite{Kavanagh07}. Characterising CH properties allows us to determine the arrival time of high speed solar wind streams at Earth and other positions in the heliosphere.

CHs are regions of the solar atmosphere that contain `open' magnetic field lines which extend to interplanetary space and give rise to high-speed solar wind streams \cite{Altschuler72,Krieger73,Hassler99,Schwadron03,Antonucci04,Fujiki05,Temmer07}. The electron density in CHs is 2-3 times lower than that of the quiet Sun (QS), while the temperature is around 10$^{5}$-10$^{6}$~K \cite{Munro72,Wiegelmann05,Wilhelm06}. As a result of their lower density and emission measure, CHs appear as dark areas in X-ray and extreme ultraviolet (EUV) wavelengths \cite{Reeves70,Vaiana76}. Magnetograms have also shown CHs to be of one predominant polarity \cite{Wiegelmann04}.

There have been a number of attempts to study CHs from a theoretical perspective. One of the main theories is the interchange model \cite{Wang96,Lionello06}, which is based on the assumption that the dominant process in the coronal `open' field evolution is the reconnection between `open' and closed magnetic field lines. This view was initiated by observations showing streamer evolution (inflows and outflows) which could be explained with the closed flux opening into the solar wind and 'open' flux closing back into the streamer \cite{Sheeley02}. Interchange reconnection could also explain the rigidly rotating CHs that are resistant to differential rotation \cite{Timothy75}. The counter-argument to interchange reconnection is that the continuous opening and closing of coronal field lines would result in the injection of magnetic flux into the solar wind and cause a bi-directional heat flux \cite{Antiochos07}. The other main theory is the quasi-steady corona-wind model in which the coronal magnetic field is extrapolated from the photosphere using magnetograms. The simplest version, the source surface model, provides us with a topology of closed and `open' flux regions by assuming a current-free corona and a completely radial magnetic field at a fixed surface \cite{Luhmann03,Schrijver03}. To obtain the evolution of the coronal field, a sequence of the source surface solutions are used. In recent years, the quasi-steady model has been improved by including the solution of the full MHD equations, which made the assumption of a fixed surface and a current-free corona unnecessary \cite{Pisanko97,Roussev03,Odstrcil03}.

To study their physical properties, CHs must be detected in a consistent manner. Initial attempts to extract coronal holes involved hand-drawn maps based on the two-day average of He~{\sc i} (10830~\AA) images and magnetograms \cite{Harvey02} made with the {\it National Solar Observatory/Kitt Peak Vacuum Telescope} (NSO/KPVT). In recent years, a number of automated selection methods have been developed: \inlinecite{Henney05} used NSO/KPVT He~{\sc i} (10830~\AA) images, magnetograms and the above-mentioned hand-drawn CH maps; \inlinecite{Malanushenko05} used He~{\sc i} imaging spectroscopy to determine CH boundaries; while \inlinecite{deToma05} have combined space-based EUV images with ground based He~{\sc i} (10830~\AA) and H$\alpha$ observations and magnetograms to identify coronal hole regions. \inlinecite{Chapman02} and \inlinecite{Chapman07} constructed synoptic CH maps using measurements made with the {\it Solar \& Heliospheric Observatory Coronal Diagnostic Spectrometer} (SOHO/CDS; \citeauthor{Harrison97}\citeyear{Harrison97}). \inlinecite{Barra07} used a multichannel segmentation method on EUV images to differentiate CHs, QS and active regions (ARs). More recently, \inlinecite{Scholl08} applied an image segmentation technique based on a combination of region and edge-based methods using 171~\AA\ (Fe~{\sc ix/x}), 195~\AA\ (Fe~{\sc xii}), 304~\AA\ (He~{\sc ii}) images and {\it Michelson Doppler Imager} (MDI; \citeauthor{Scherrer95}\citeyear{Scherrer95}) magnetograms to determine CH boundaries.

Although determining the intensity threshold is of fundamental importance to accurately determine CH boundaries (and thus their properties), many of the methods discussed rely on arbitrary choices to determine the CH threshold. The main goal of our study was to produce a fast and robust algorithm to automatically identify coronal holes at any time of the solar cycle using a single image (195~\AA\ or X-ray) at a time. We introduced a three perspective identification of CHs using the {\it Solar Terrestrial Relations Observatory/Extreme Ultraviolet Imager} (STEREO-SECCHI/EUVI; \citeauthor{Wuelser04}\citeyear{Wuelser04}) and the SOHO {\it Extreme Ultraviolet Imaging Telescope} (EIT; \citeauthor{Delaboudiniere95}\citeyear{Delaboudiniere95}). The properties of several CHs were then compared with solar wind speed measurements made by the STEREO {\it Plasma and Suprathermal Ion Composition instrument} (PLASTIC; \citeauthor{Blush05}\citeyear{Blush05}) and the {\it Advanced Composition Explorer} (ACE; \citeauthor{Chiu98}\citeyear{Chiu98}) to study the connection between the areas of CHs and solar wind flows at a number of perspectives at 1~AU.

\section{Observations}

The segmentation algorithm was tested on 195~\AA\ images made with the STEREO-SECCHI/EUVI and SOHO/EIT instruments and X-ray images made with the {\it Hinode X-Ray Telescope} (XRT; \citeauthor{Golub07}\citeyear{Golub07}; \citeauthor{Kano08}\citeyear{Kano08}). The 195~\AA\ observations show the inner solar corona with a difference in the distribution of intensities between coronal holes and the quiet Sun. In this study, full resolution 1024 $\times$1024 (EIT) and 2048 $\times$ 2048 pixel (EUVI and XRT) images were used. EIT 195~\AA\ images are normally produced every 12 minutes, and EUVI images every 10 minutes. The 195~\AA\ filter has a broad temperature response, which peaks at approximately 1.5$\times$10$^{6}$ K. 1024 $\times$1024 pixel images made by XRT were also used in this study. The wavelength range of the XRT instrument is 6-60~\AA, the temperature response of the Ti-Poly filter (27.4~\AA) used in this study peaks at around 9$\times$10$^{6}$ K. MDI line-of-sight magnetograms were used to differentiate coronal holes from filaments. 

The threshold of one large CH visible with the four satellites around 12-15 January 2008 was studied in detail, while the long-term threshold variation was shown using observations obtained over 30 days in January 2008. We have also investigated the different line-of-sight magnetic field ($B_{LOS}$) distributions for low intensity regions (LIRs) observed on 15 June 2008 and 22 March 2007. Finally, the characteristics of 46 large CHs were studied for a three-year period starting in 2006. Only major CHs with distinct solar wind profiles were included in the analysis to compare their area with high-speed solar wind stream durations.

\section{Data Analysis}

\subsection{Local Intensity Thresholding}

Standard {\it SolarSoft} procedures \cite{Freeland98} were used to reduce and calibrate the images, which were then transformed from full-disk images to Lambert equal-area projection (LEAP) maps. The latter procedure projects a spherical surface onto a cylindrical surface whilst preserving the area and producing parallel latitudinal (and longitudinal) lines within approximately -60 and +60 degrees latitude and longitude \cite{McAteer05}. The coordinates of the coronal hole boundaries are only reliable within the above mentioned range due to near-limb projection effects. 

\begin{figure} 
\centerline{\hspace*{0.015\textwidth} 
\includegraphics[angle=90,clip=, width=1\textwidth]{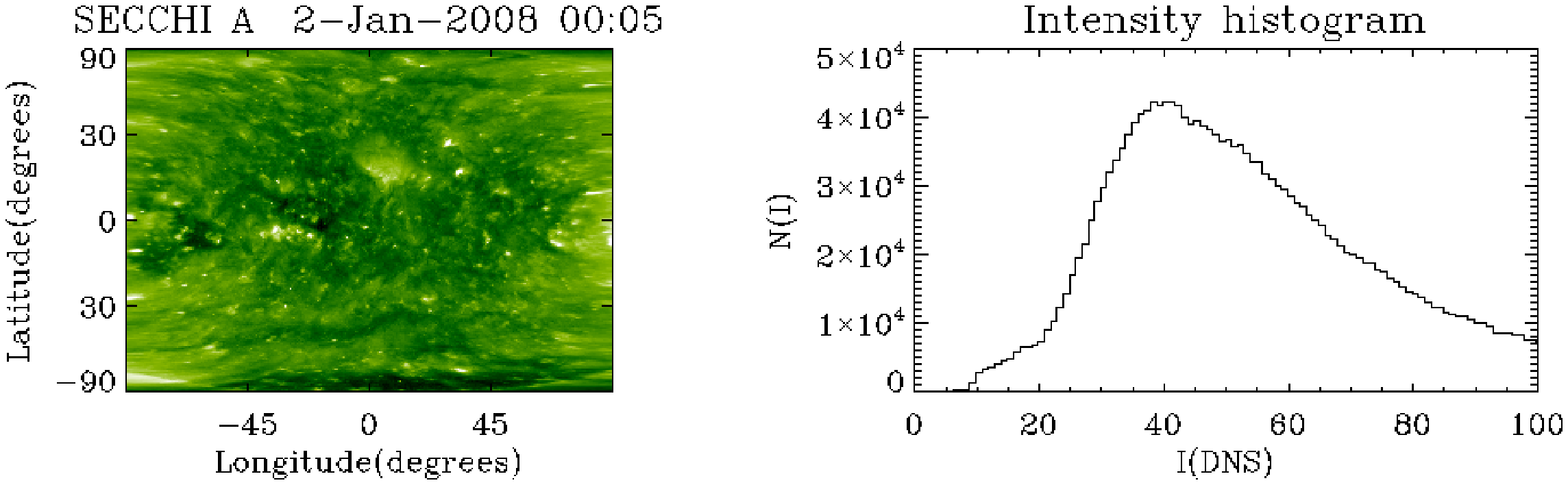} } 
\vspace{-0.35\textwidth} 
\centerline{\Large \bf     
\hfill} 
\vspace{0.31\textwidth}    
\centerline{\hspace*{0.015\textwidth} 
\includegraphics[angle=90,clip=, width=1\textwidth]{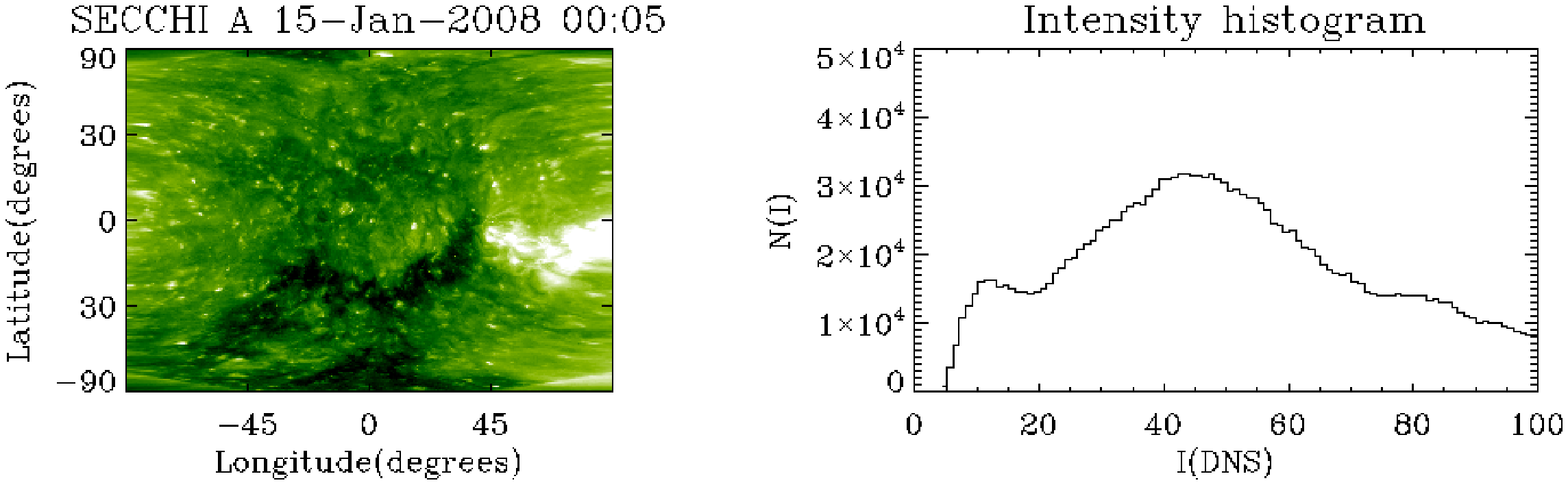} } 
\caption{{\it Top}: EUVI 195~\AA\ LEAP map showing no CHs, and the corresponding intensity histogram with a unimodal distribution. {\it Bottom:} LEAP map showing a large CH, and the corresponding intensity histogram with a bimodal distribution.} 
\label{carr_maps} 
\end{figure}

Intensity thresholding is a powerful method for identifying features in images. It is well known that the intensity histogram of an image gives a multimodal distribution where each frequency distribution corresponds to a feature in the image \cite{Gonzalez02,Gallagher98,Gallagher99,Belkasim03}. In the case of 195~\AA\ images, each major feature (LIRs, QS and ARs) corresponds to a distribution in the intensity histogram. When an image is dominated by quiet Sun, the corresponding intensity histogram shows a unimodal distribution (top panel of Figure~\ref{carr_maps}). When a relatively large part of the image is covered by LIRs, the resulting intensity histogram shows a bimodal distribution (bottom panel of Figure~\ref{carr_maps}), where the local minimum corresponds to the boundary threshold. 

\begin{figure} 
\centerline{\hspace*{0.015\textwidth} 
\fbox{ 
\includegraphics[clip=, width=7cm]{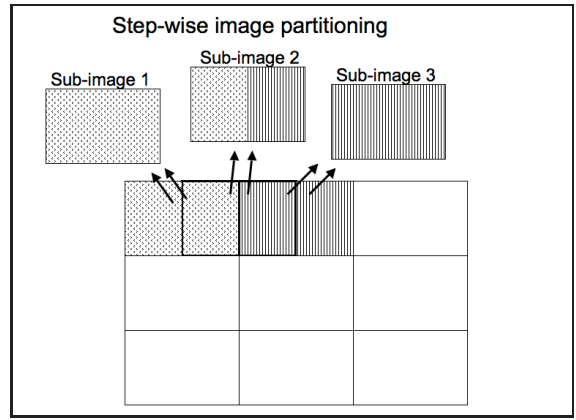} }} 
\caption{The LEAP map is partitioned into overlapping sub-images.} 
\label{windowing} 
\end{figure}

\begin{figure} 
\centerline{\hspace*{0.015\textwidth}
\includegraphics[angle=90,clip=, width=10cm]{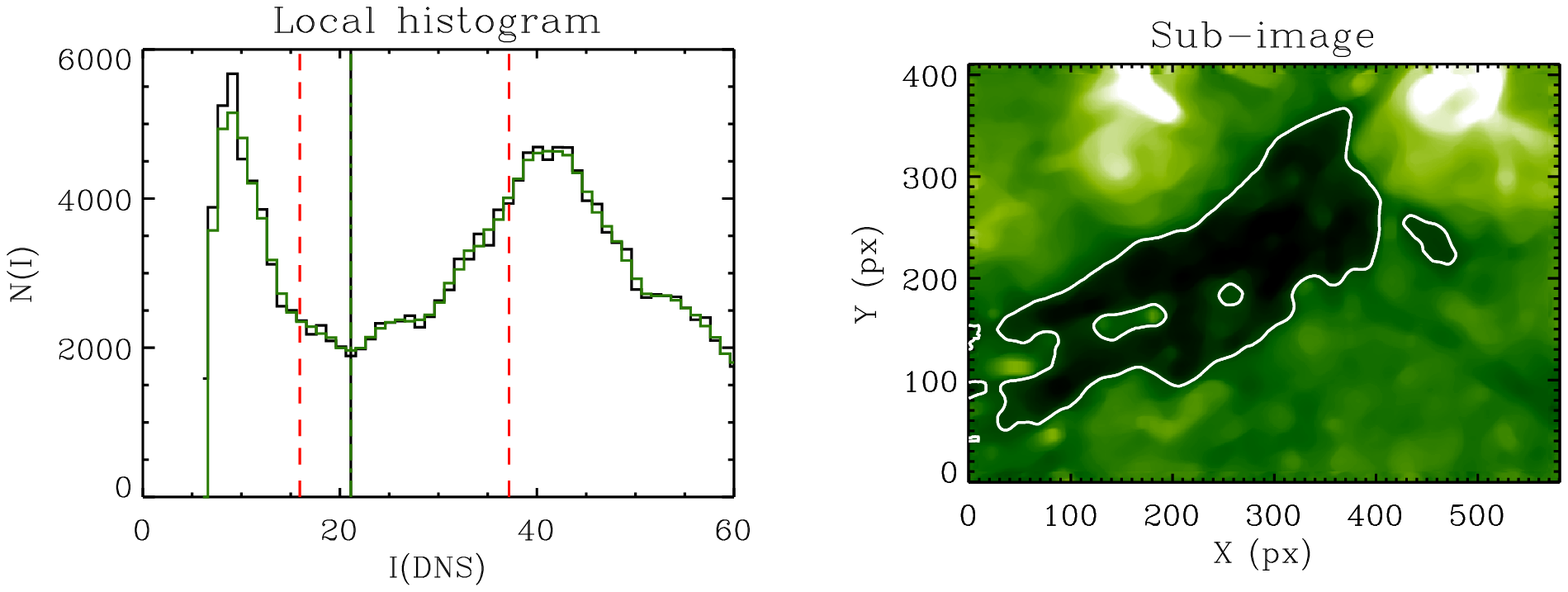} } 
\vspace{-0.35\textwidth}   
\centerline{\Large \bf 
\hfill} 
\vspace{0.31\textwidth}    
\centerline{\hspace*{0.015\textwidth} 
\includegraphics[angle=90,clip=, width=10cm]{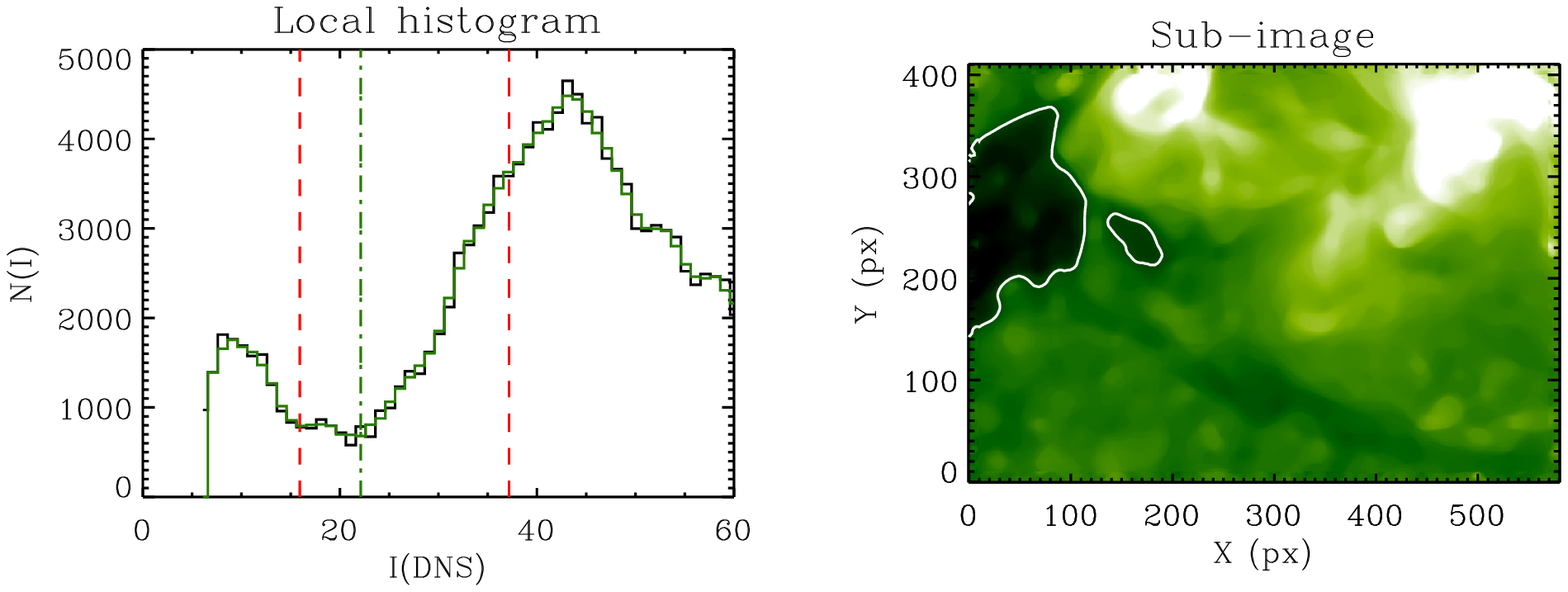} } 
\caption{{\it Right}: The corresponding local histograms with minimum positions marked with dot-dashed line. The range (30--70\% of QS intensity) where the minimum can be found is marked with dashed lines. {\it Left}: Sub-images from a EUVI 195~\AA\ LEAP map (11 January 2008).} 
\label{subimages} 
\end{figure}

The minimum between the LIR and QS distribution was enhanced using a partitioning operation. By partitioning the solar image to sub-images, we obtained local histograms with more defined minima which aided the determination of thresholds. The sub-images used were 1/2, 1/3, 1/4 and 1/5 of the original LEAP map size. Each of the sub-image sizes were used to step across the map and obtain local intensity histograms from overlapping sub-images (Figure~\ref{windowing}). This process provided more clearly discernible local minima for thresholding (Figure~\ref{subimages}). This can be explained by different size sub-images having different proportions of QS and LIR and hence frequency distributions. The threshold minima are defined as local minima that are located within 30--70\% of the QS intensity and are wider than 6 digital numbers (DNS).

As mentioned above, the intensity range where the thresholds were identified depends on the QS intensity. For this reason the method is reliable at any time of the solar cycle regardless of the change in the overall intensity. It should be noted though, that with increasing solar activity, the acquired CH boundaries are less accurate due to bright coronal loops intercepting the line-of-sight and obscuring parts of the CH boundary.

\begin{figure} 
\centerline{\hspace*{0.015\textwidth}
\includegraphics[angle=90,trim=0 0 0 50, width=12cm]{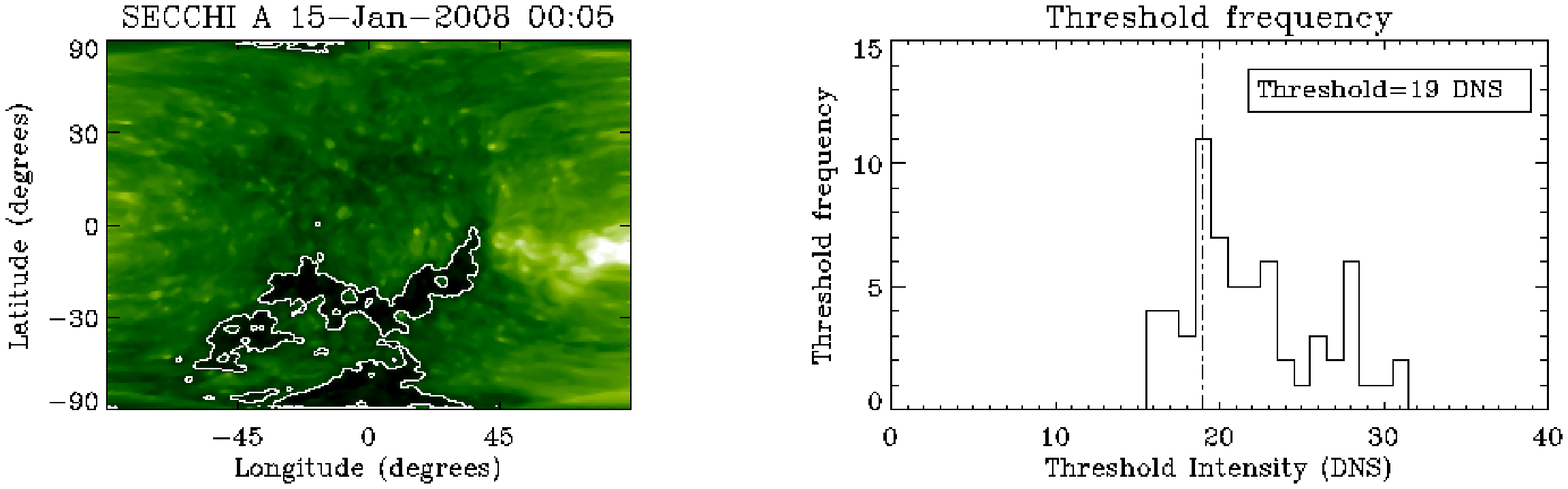} } 
\vspace{-0.35\textwidth}   
\centerline{\Large \bf 
\hfill} 
\vspace{0.31\textwidth} 
 
\centerline{\hspace*{0.015\textwidth} 
\includegraphics[angle=90,trim=0 0 0 50, width=12cm]{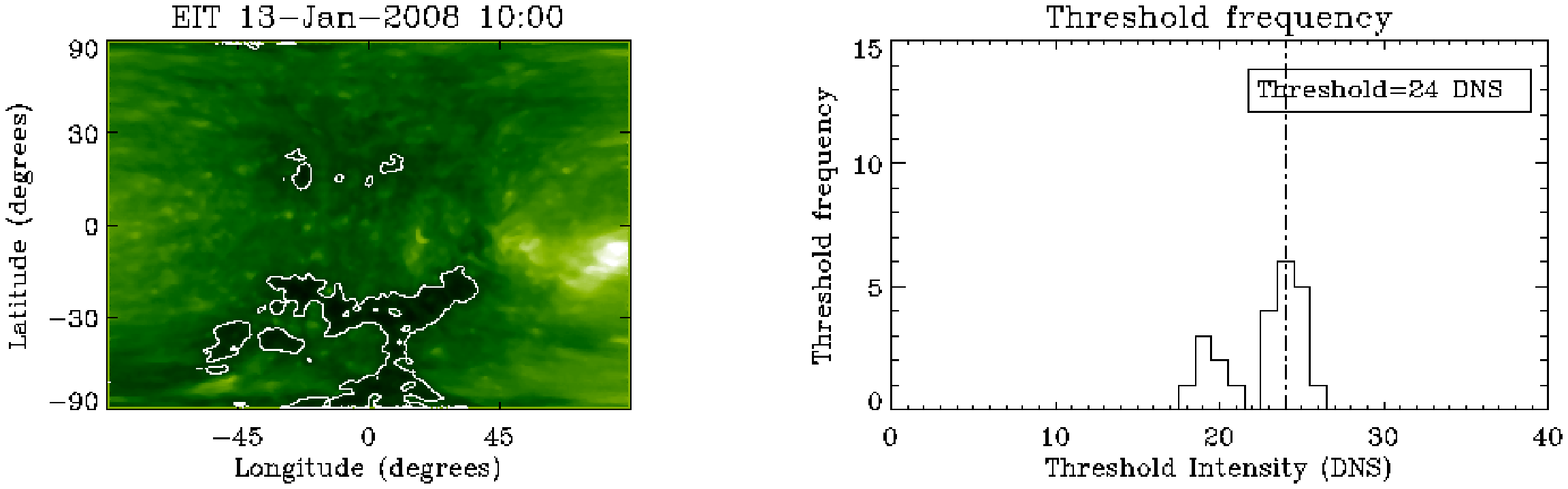} } 
\vspace{-0.35\textwidth}   
\centerline{\Large \bf     
\hfill} 
\vspace{0.31\textwidth} 

\centerline{\hspace*{0.015\textwidth}  
\includegraphics[angle=90,trim=0 0 0 50, width=12cm]{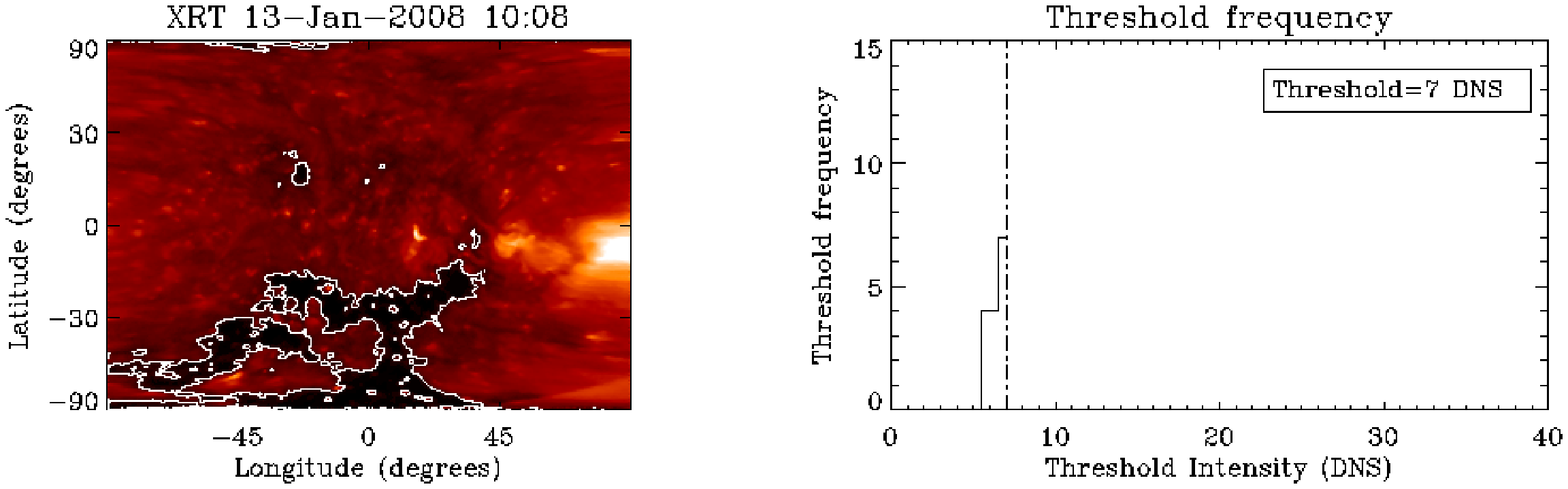} } 
\vspace{-0.35\textwidth}   
\centerline{\Large \bf 
\hfill} 
\vspace{0.31\textwidth}
 
\centerline{\hspace*{0.015\textwidth} 
\includegraphics[angle=90,trim=0 0 0 50, width=12cm]{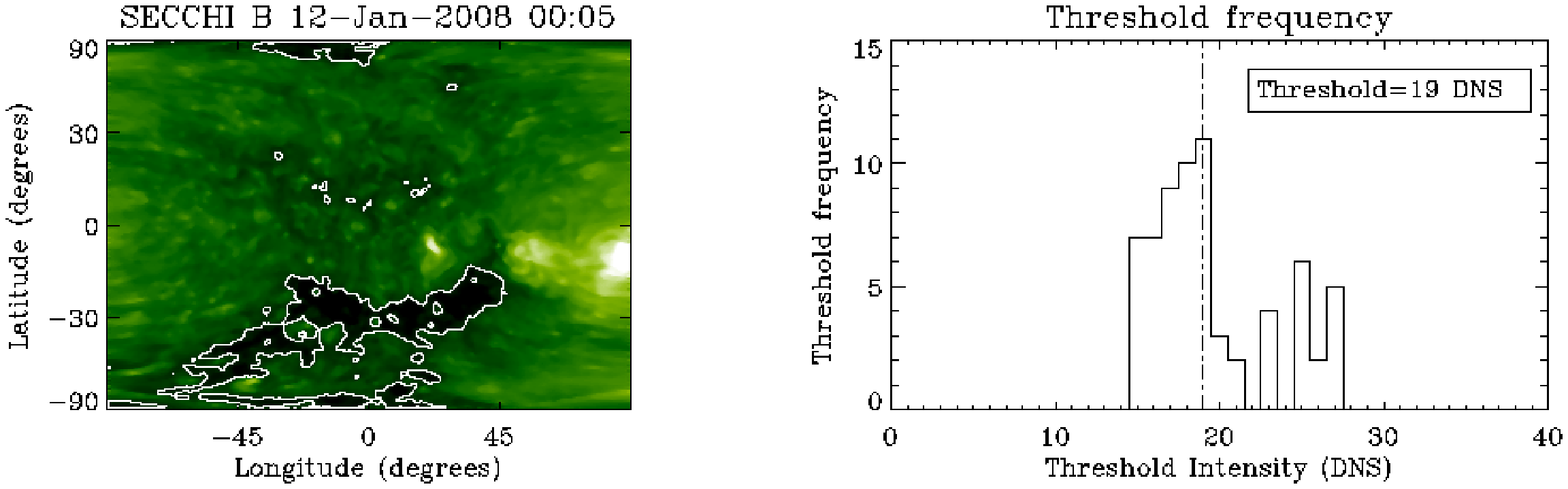} }
\caption{{\it Left}: STEREO A/EUVI, SOHO/EIT, Hinode/XRT and STEREO B/EUVI LEAP maps with LIR contours. The images show a complex CH structure approximately at the same location on the map. {\it Right}: Threshold histogram, where the peak value is the optimal threshold.} 
\label{lambertmaps} 
\end{figure}

Once the final threshold was obtained, the LIR maps were created. Figure~\ref{lambertmaps} shows the same CH complex observed by four different instruments in two wavebands (195~\AA\ and X-ray). The observation times are different to ensure that the CH is seen at the same central location observing with different satellites. We would like to draw attention to the large-scale similarity in CH morphology despite the use of different instruments and observation times. Also, the threshold histograms (right column of Figure~\ref{lambertmaps}) show how distinct the peak threshold is, which is then used to contour the LIRs.
 
\subsection{Differentiating Coronal Holes from Filaments}

\begin{figure} 
\centerline{\hspace*{0.015\textwidth} 
\includegraphics[angle=90,clip=true,trim= 0 25 0 0, width=14cm]{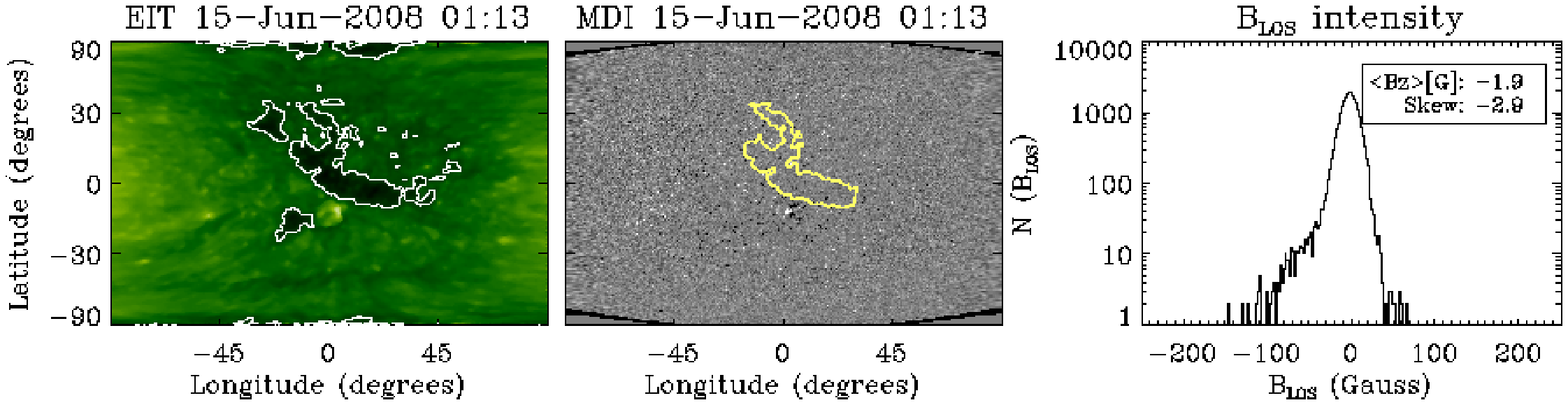} }
\vspace{-0.35\textwidth}   
\centerline{\Large \bf     
\hfill} 
\vspace{0.21\textwidth} 

\centerline{\hspace*{0.015\textwidth} 
\includegraphics[angle=90,clip=true,trim= 0 25 0 0, width=14cm]{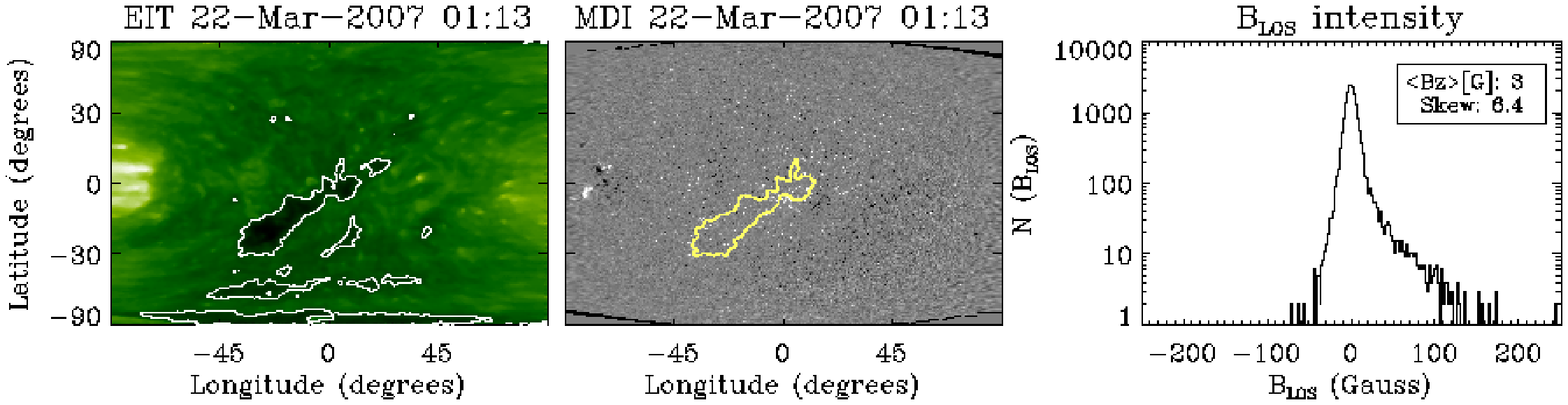} }
\vspace{-0.35\textwidth}   
\centerline{\Large \bf     
\hfill} 
\vspace{0.21\textwidth} 

\centerline{\hspace*{0.015\textwidth} 
\includegraphics[angle=90,clip=true,trim= 25 25 0 0, width=14cm]{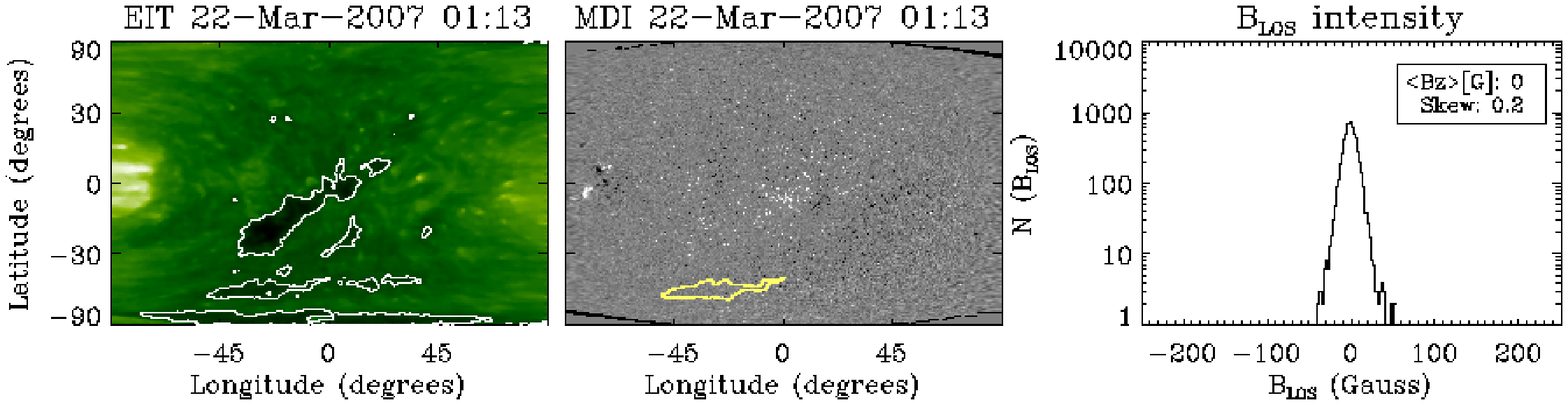} } 
\caption{{\it Left}: EIT LEAP maps with LIR contours. Middle: MDI LEAP maps with LIR contours overlaid. {\it Right}: $B_{LOS}$ intensity histogram of the contoured regions. Regions highlighted in the first and second middle panels are CHs (B$_{LOS}$ histograms skewed), the third middle panel shows a filament ($B_{LOS}$ histogram not skewed).}
\label{mdi} 
\end{figure}

CHs were distinguished from filaments using MDI magnetograms. A magnetogram was found closest in time to the EUV observation and then differentially rotated to the EUV observation time. The LIR contours obtained from the EUV image were then overplotted on the differentially rotated magnetogram and the flux imbalance was determined for the contoured regions. CHs are well know to show one dominant polarity, while filaments have a bi-polar distribution of polarities \cite{Scholl08}. Figure~\ref{mdi} shows examples of LIRs observed on 15 June 2008 and 22 March 2007 respectively. The 15 June 2008 LIR is an example for a CH with a negative dominant polarity, while the 22 March 2007 LIR is a CH with a positive dominant polarity (first and second row in Figure~\ref{mdi}). The skewness of the field strength distribution was found to be -2.9 and 6.4 respectively for the two CHs. The LIR highlighted on the bottom panel magnetogram in Figure~\ref{mdi} was found to be a filament based on the very small value of skewness (0.2) in its line of sight magnetic field strength distribution.

\section{Results}

\begin{figure} 
\centerline{\hspace*{0.015\textwidth} 
\includegraphics[clip=, width=12cm]{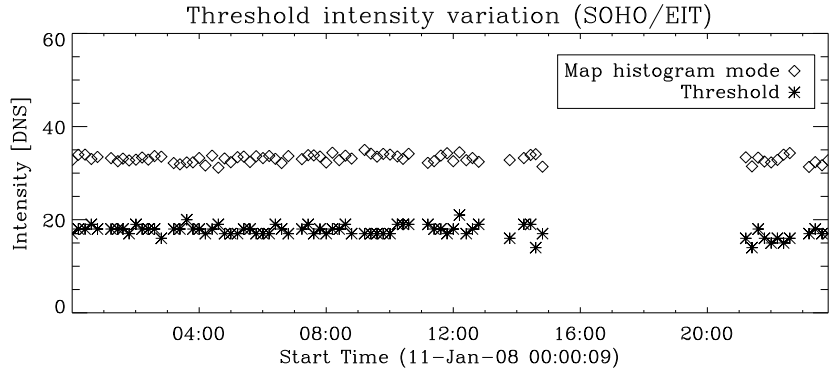} } 
\caption{The threshold values obtained from 76 EIT 195~\AA\ images taken on 11 January 2008. The gap in the data between 15:00 to 21:00 is due to a data drop-out.} 
\label{ch_day} 
\end{figure}

\begin{figure} 
\centerline{\hspace*{0.015\textwidth} 
\includegraphics[clip=true,trim= 0 35 0 0, width=\linewidth]{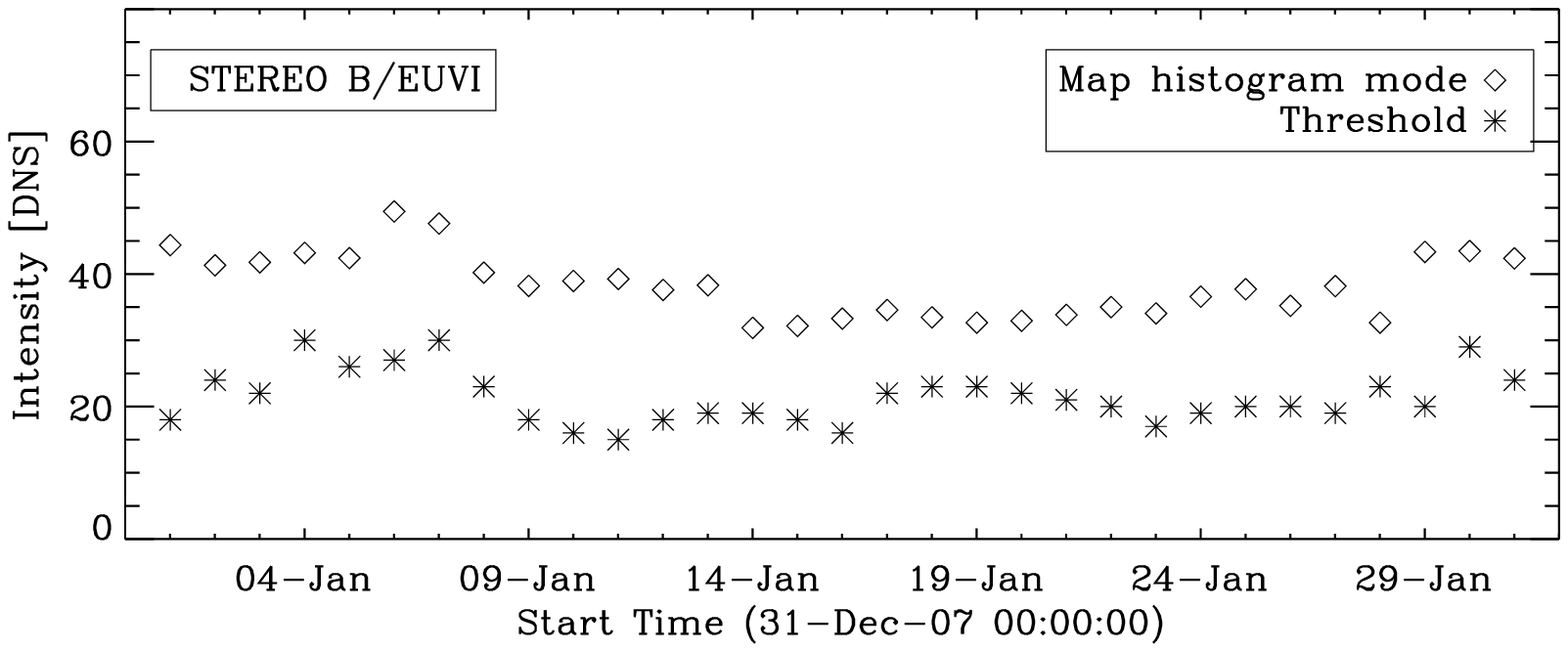} } 
\vspace{-0.35\textwidth}   
\centerline{\Large \bf     
\hfill} 
\vspace{0.263\textwidth}    

\centerline{\hspace*{0.015\textwidth} 
\includegraphics[clip=true,trim= 0 35 0 0, width=\linewidth]{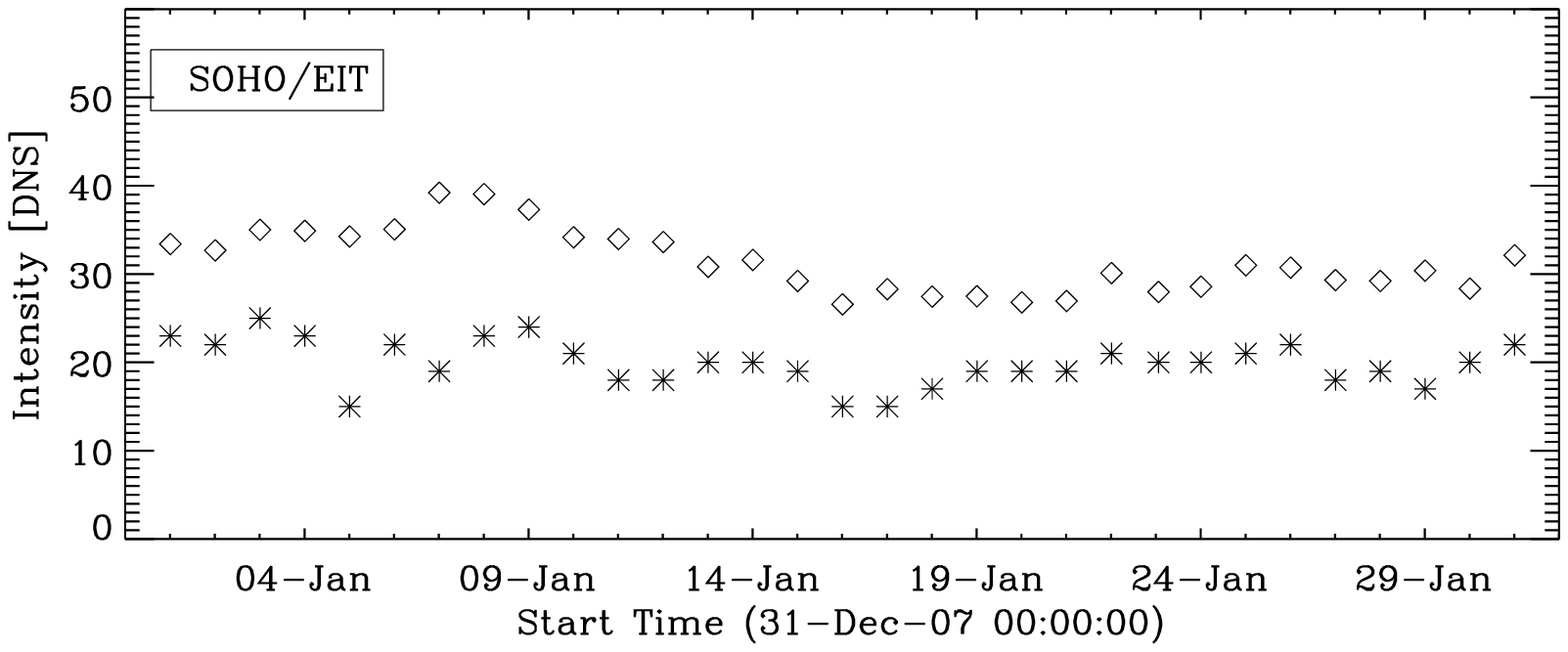} } 
\vspace{-0.35\textwidth}   
\centerline{\Large \bf     
\hfill} 
\vspace{0.263\textwidth}    

\centerline{\hspace*{0.015\textwidth} 
\includegraphics[clip=true,trim= 0 35 0 0, width=\linewidth]{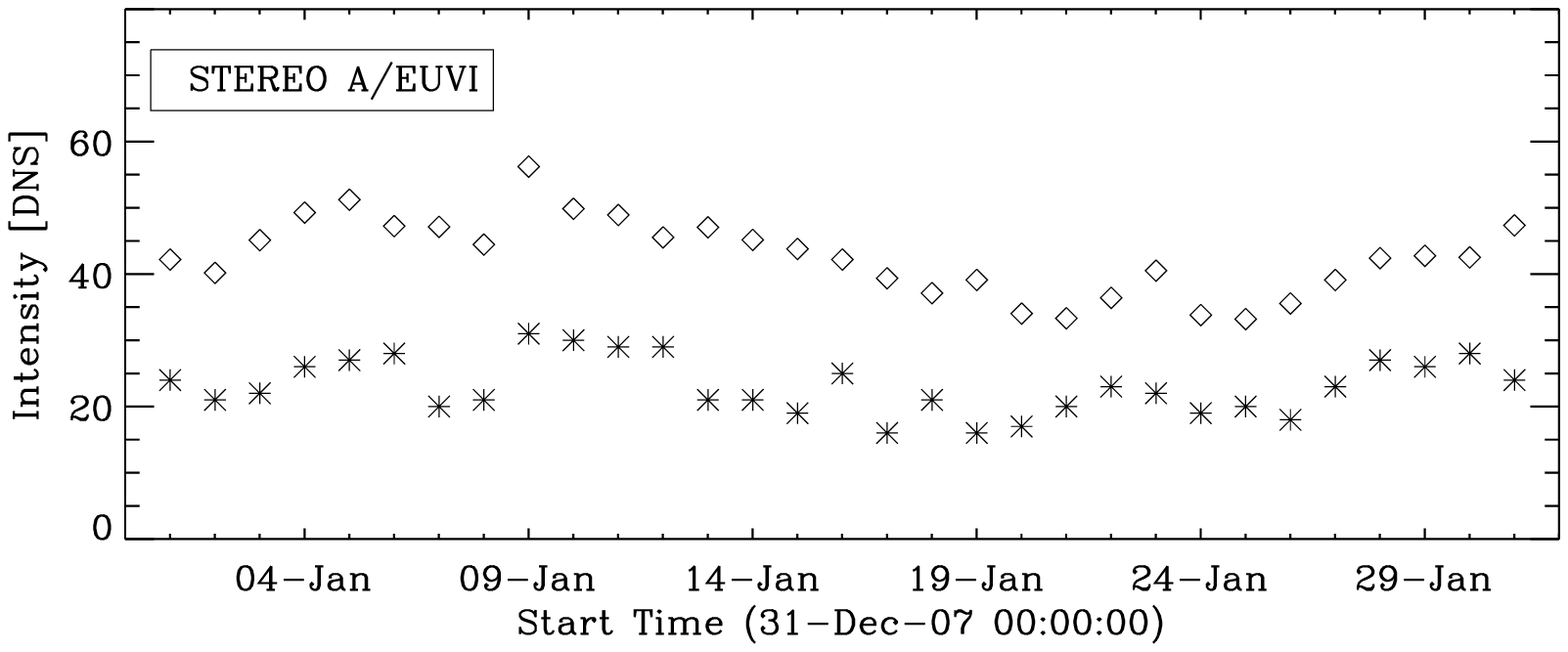} } 
\vspace{-0.35\textwidth}   
\centerline{\Large \bf     
\hfill} 
\vspace{0.263\textwidth}    

\centerline{\hspace*{0.015\textwidth} 	
\includegraphics[clip=, width=\linewidth]{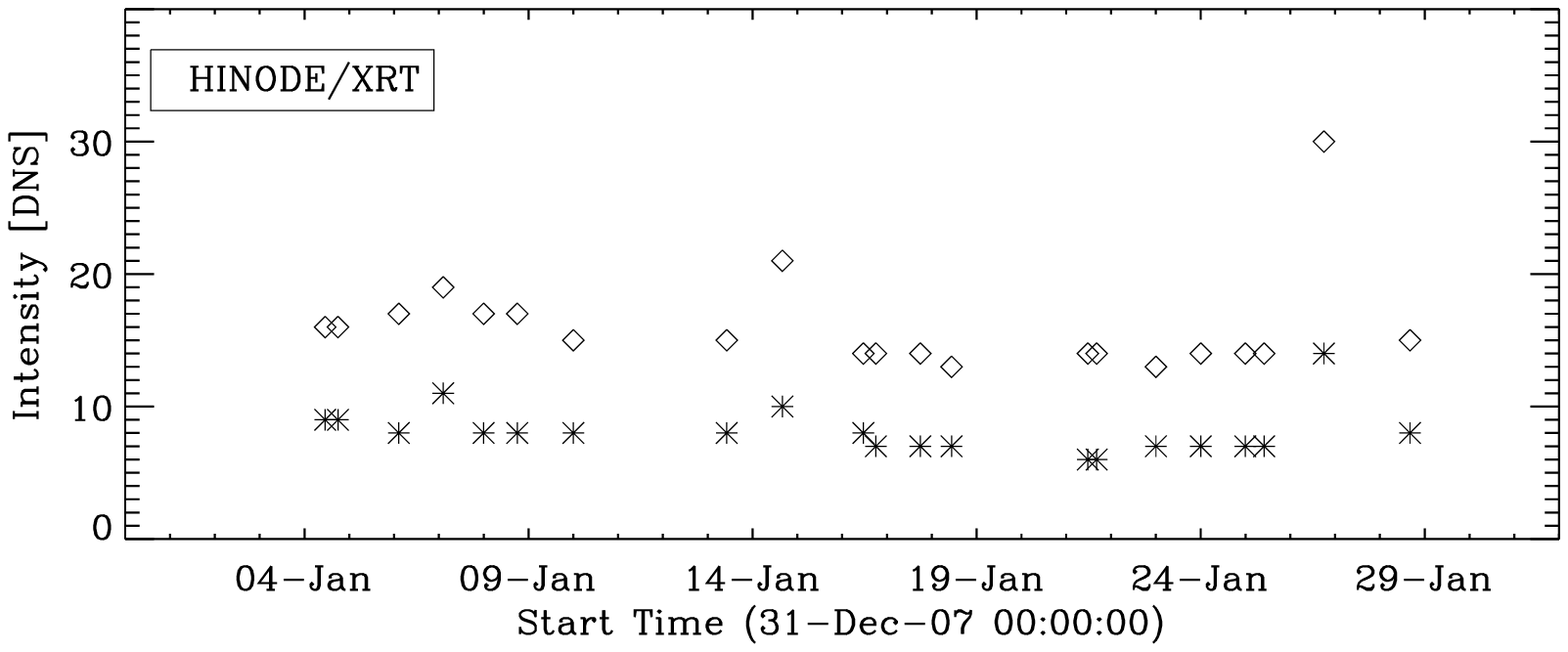} } 
\caption{Variation of the LIR threshold intensity over a month (January 2008) obtained using STEREO/EUVI B, SOHO/EIT, STEREO/EUVI A 195~\AA\ and Hinode/XRT X-ray images. } 
\label{ch_month} 
\end{figure}

Figure~\ref{ch_day} shows a threshold consistency test. In this example, thresholds were acquired from 76 EIT 195~\AA\ images taken on 11 January 2008. It was expected that thresholds of similar intensity would be found due to the similarity of the images taken close in time (images were taken approximately every 12 minutes). The mean intensity threshold was found to be (17.6$\pm$1.2)~DNS. Assuming that the intensity threshold distribution is normal, the true intensity threshold falls within $\pm$14\% of the calculated mean intensity threshold to 95\% confidence. This confirms that the calculated thresholds are stable. Figure~\ref{ch_month} shows the variation of the LIR threshold over an entire month (January 2008) using four instruments onboard STEREO, SOHO and Hinode.

\begin{figure} 
\centerline{\hspace*{0.015\textwidth} 
\includegraphics[angle=90,clip=, width=14cm]{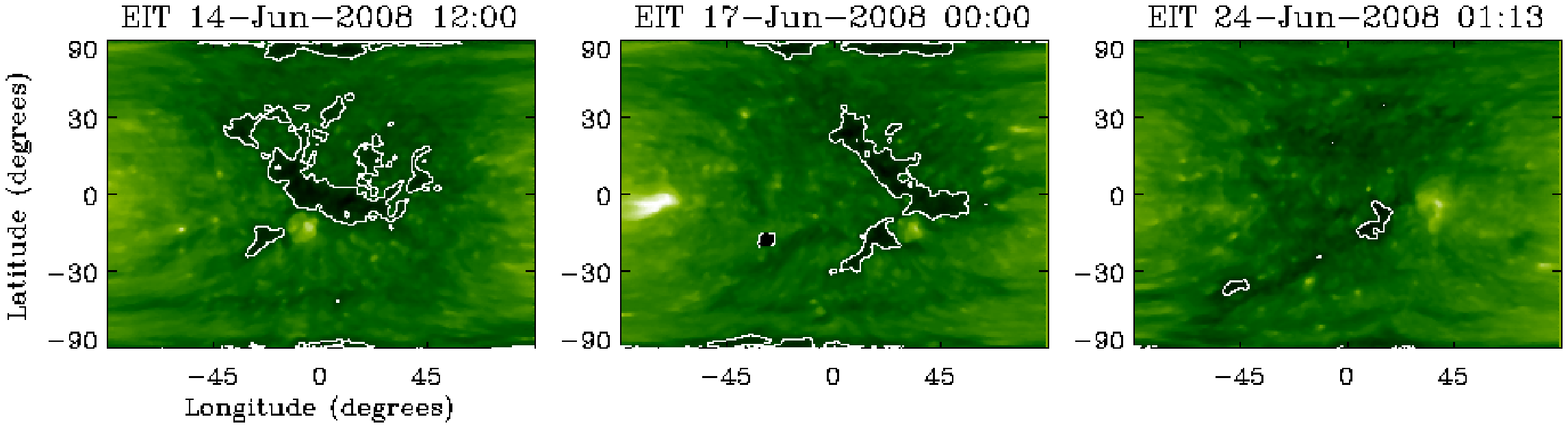} } 
\vspace{-0.40\textwidth}   
\centerline{\Large \bf 
\hfill} 
\vspace{0.3\textwidth} 

\centerline{\hspace*{0.015\textwidth} 
\includegraphics[angle=90, clip=, width=12cm]{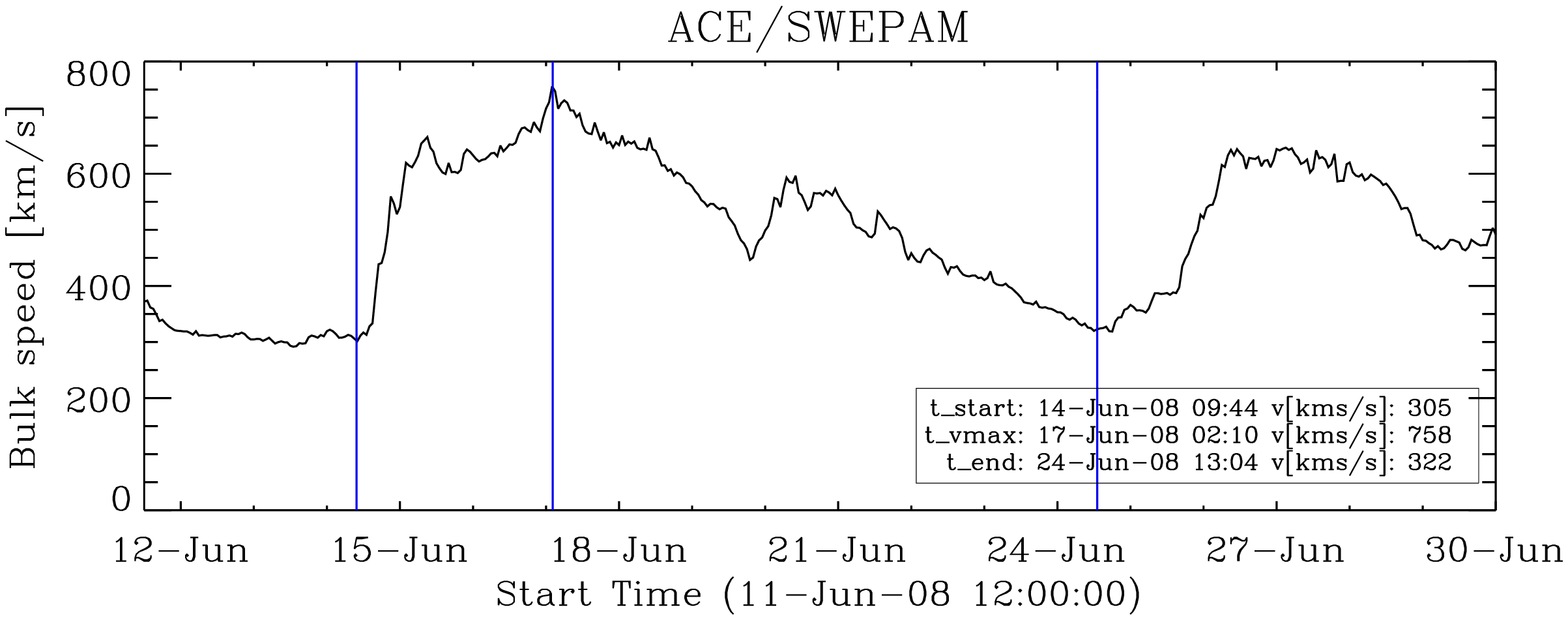} } 
\caption{ {\it Top}: EIT LEAP maps showing the CH position on the day of the start, the maximum and the end of the high-speed solar wind period. {\it Bottom}: The corresponding ACE solar wind speed measurements for the studied period.} 
\label{ace} 
\end{figure}

\begin{figure} 
\centerline{\hspace*{0.015\textwidth} 
\includegraphics[angle=90,clip=, width=13cm]{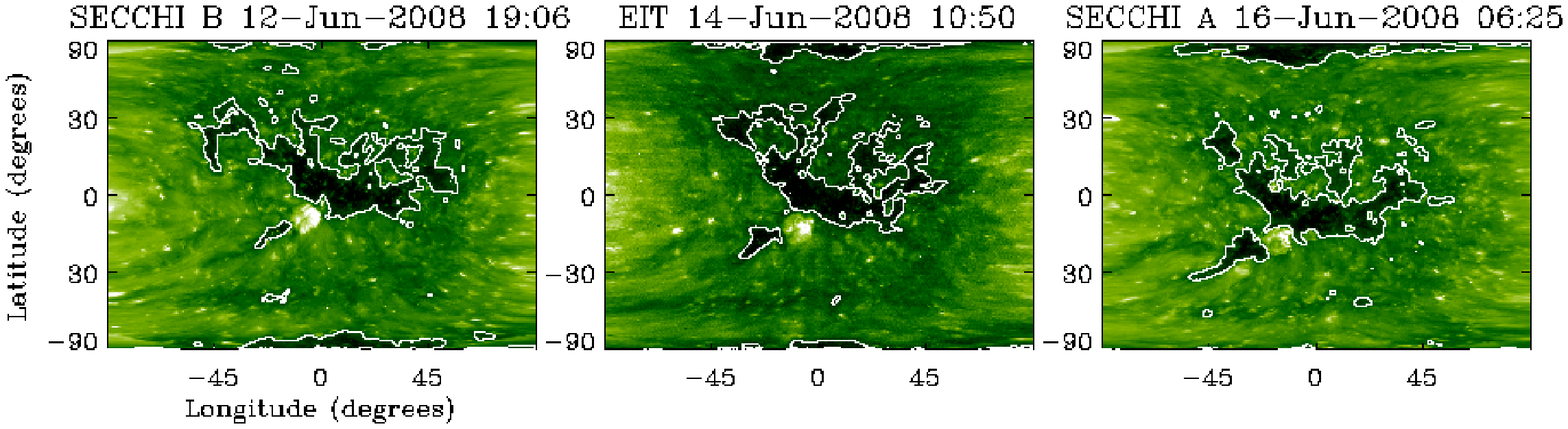} } 
\vspace{-0.35\textwidth}   
\centerline{\Large \bf 
\hfill} 
\vspace{0.27\textwidth} 

\centerline{\hspace*{0.015\textwidth} 
\includegraphics[angle=90,clip=true,trim= 20 0 0 0, width=11cm]{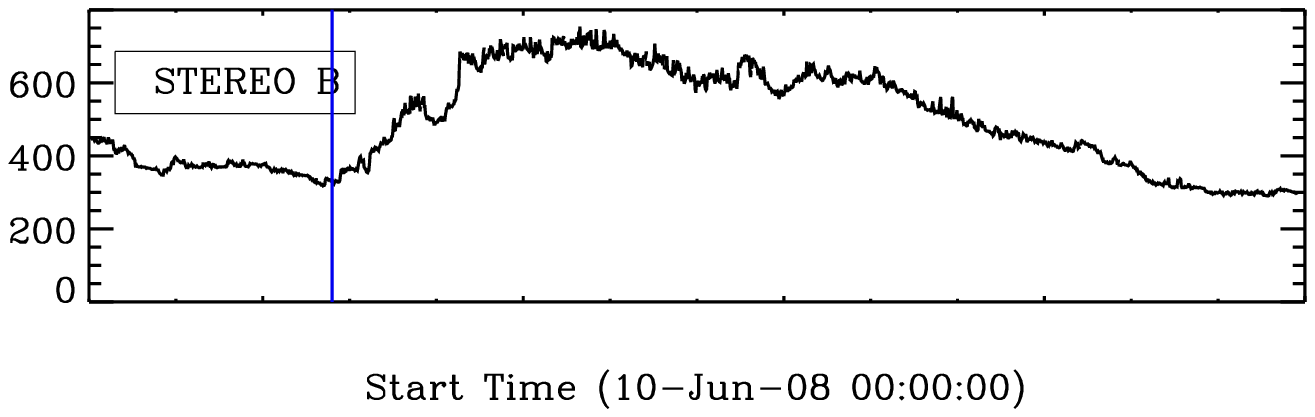} } 
\vspace{-0.35\textwidth}   
\centerline{\Large \bf 
\hfill} 
\vspace{0.229\textwidth} 

\centerline{\hspace*{0.015\textwidth} 
\includegraphics[angle=90,clip=true,trim= 20 0 0 0, width=11cm]{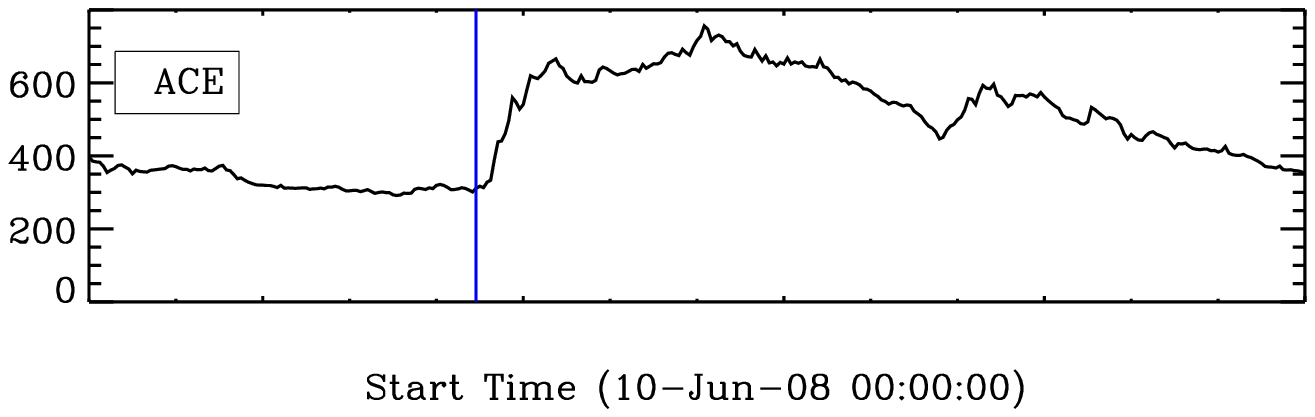} } 
\vspace{-0.35\textwidth}   
\centerline{\Large \bf 
\hfill} 
\vspace{0.229\textwidth} 

\centerline{\hspace*{0.015\textwidth} 
\includegraphics[angle=90,clip=, width=11cm]{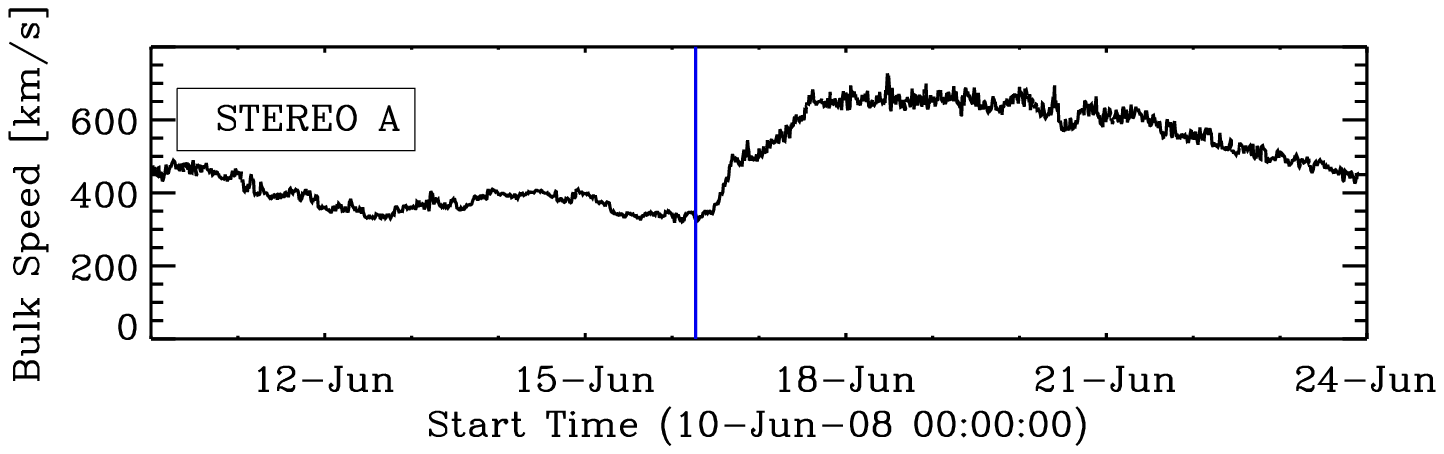} } 

\caption{ {\it Top}: EIT LEAP maps showing the CH position on three different days when the high-speed solar wind reached STEREO B, ACE and STEREO A. {\it Bottom}: The solar wind profiles obtained from the three satellites, with the line marking the start of the high-speed solar wind.}
\label{sw_profiles} 
\end{figure}

All coronal holes studied here were linked to significant rises in solar wind speeds at 1~AU. Figure~\ref{ace} shows the CH positions at the start, maximum and end of the high-speed solar wind period on 14--24 June 2008. The images show that the west-most CH boundary was at approximately W50  when the high-speed solar wind stream arrived at ACE. The CH has passed over the west limb by the time the solar wind speed dropped back to normal levels (300--400~km~s$^{-1}$). The duration of the high-speed solar wind stream is related to the expansion factor of the magnetic flux-tube rooted in the CH. This is discussed in more detail below.

The relationship between this CH and its associated solar wind stream measured at three distinct points in the heliosphere was explored using imaging and {\it in-situ} measurements from STEREO B, ACE and STEREO A. The top row of Figure~\ref{sw_profiles} shows the CH complex on three days during 14-24 June 2008. The corresponding solar wind velocity measurements, given in the bottom three panels of Figure~\ref{sw_profiles}, show a high-speed solar wind stream sweeping around the inner heliosphere, from STEREO B to ACE to STEREO A. The high-speed stream was initially detected by STEREO B at 12 June 2008 19:10~UT, when the west-most boundary of the CH was at $\sim$W53 from the perspective of STEREO B. Subsequently, the solar wind speed began to rise in ACE {\it in-situ} measurements two days later on 14 June 2008 11:00~UT. At this point in time, the CH's west-most boundary was at $\sim$W50 as seen by SOHO. The high-speed stream finally traversed the point in the heliosphere monitored by STEREO A on 16 June 2008 6:30~UT, at which time the western boundary of the CH was at $\sim$W43. The complementary imaging and {\it in-situ} capabilities of SOHO, ACE and the two STEREO satellites offer a unique opportunity to study the relationship between CHs rotating across the solar disk and high-speed solar wind streams sweeping through the heliosphere.
 
\begin{figure} 
\centerline{\hspace*{0.015\textwidth} 
\includegraphics[clip=, width=12cm]{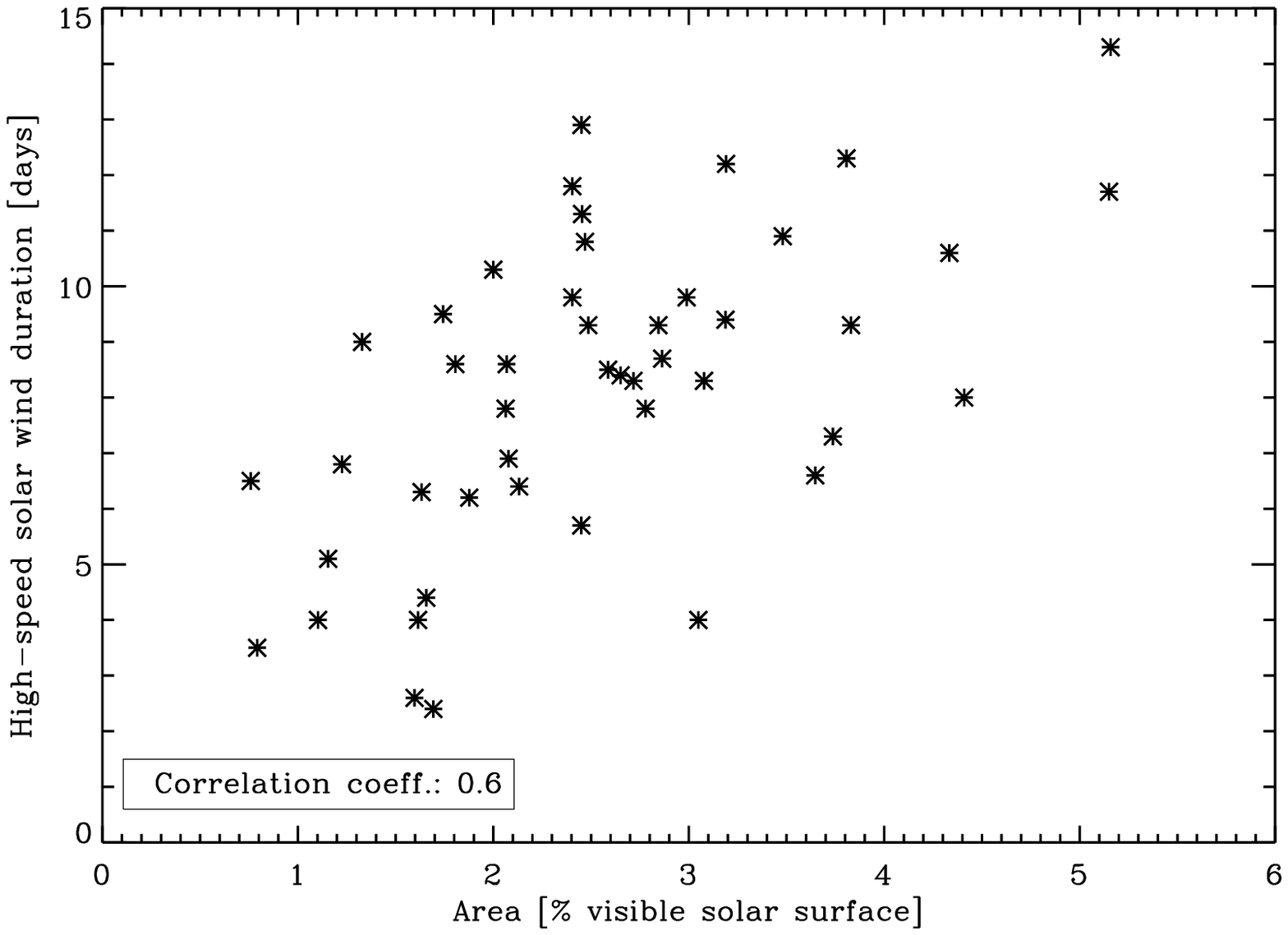} } 
\vspace{-0.35\textwidth}   
\centerline{\Large \bf 
\hfill} 
\vspace{0.31\textwidth} 
\caption{Scatterplot showing the relationship between the area of CHs and the high-speed solar wind duration caused by the corresponding CHs. 46 CHs were included in this study.}
\label{scatterplot} 
\end{figure}

As indicated previously, it is well-known that there is a relationship between the size of CHs and the duration of their resulting high-speed solar wind streams, whereby larger CHs produce longer duration high-speed streams. We investigated this relationship using 46 CHs which were automatically identified using our segmentation algorithm. Figure~\ref{scatterplot} shows the resulting plot of high-speed solar wind stream duration as a function of CH area. The correlation coefficient obtained was $\sim$0.6, which indicates a relationship between a CH area and the duration of a corresponding high-speed solar wind stream.

\section{Discussion and Conclusions}

We have developed a fast and robust CH identification technique that can be used with a single 195~\AA\ or X-ray image, in conjunction with a corresponding magnetogram. The segmentation algorithm first identifies LIRs using the local histograms of either 195~\AA\ or X-ray observations, and then exploits MDI magnetograms to differentiate CHs from filaments by the flux imbalance determined for the identified regions. This was achieved assuming that CHs have a predominant polarity as opposed to filaments. Once the CH regions were determined, the position, area and mean $B_{LOS}$ were calculated for multiple CHs selected in the period 2006--2008.

The main limitation of the method results from near-limb effects which include: line-of-sight obscuration in EUV and X-ray images; increase of noise in $B_{LOS}$ measurements, and interpolation effects caused by the Lambert equal-area projection. These near-limb effects can cause incorrect boundary determination, the inability to determine a dominant polarity and incorrect area determination. Also, some CH boundaries located near the limb were found to have different threshold intensities which depended on the degree to which they were obscured. For this reason, only CHs within -60 and +60 degrees latitude and longitude were included in this analysis. 

From a scientific perspective, the main goal of this study was to link the on-disk properties of CHs to high-speed solar wind streams measured with ACE and STEREO near 1~AU. The start, maximum and end of a high-speed solar wind stream was determined from ACE and STEREO solar wind measurements and the corresponding CH locations were shown on EIT and EUVI 195~\AA\ images. Figures~\ref{ace} and \ref{scatterplot} demonstrated the obvious signs of non-radial CH flux tube expansion. It must be noted that the time for passage through a solar wind stream is likely to be extended due to the dragging of the solar wind into the non-radial shape of the Parker spiral. We are currently extending this study to determine the expansion factor of CH flux tubes, and as a first step, have compared the area of 46 CHs to the corresponding high-speed solar wind periods. Although a large scatter was found in the data, a positive correlation was found (Figure~\ref{scatterplot}). Initial estimates indicate expansion factors of approximately 2--10 for flux tubes extending from the Sun to 1 AU. This assumes that the duration of a high-speed solar wind stream goes as $\tau_{duration} \sim f (25.5 CH_{width}/360)$, where $f$ is the flux-tube expansion factor and $CH_{width}$ is the CH longitudinal width. This super-radial expansion of flux-tubes rooted in CHs is consistent with previous theoretical and observational results \cite{Wang_Sheeley90,Guhathakurta98,Deforest01}.

In addition to CH flux tube expansion factors it is also important to determine {\it when} and {\it where} a particular portion of a solar wind stream emerged from the Sun. When attempting to determine {\it when} a high-speed solar wind stream will impact the Earth, the on-disk origin of the stream must be known to relatively high precision. Knowing the exact location of the CH when the high-speed plasma left, allows the determination of the path-length of the plasma. This provides the opportunity to study the modification of the Parker spiral through slow and high-speed solar wind interaction regions. Ultimately these methods will enable us to make more accurate estimates of the arrival times and durations of high-speed solar wind streams at the Earth and/or other points in the heliosphere.

We conclude that an extensive study of the connection between the CH morphology, physical properties and the solar wind profile is needed to have a better understanding of coronal holes and their interplanetary effects. Our future plans are to extend our analysis for a solar cycle and link all the major CHs to solar wind speed rises. This will allow us to study the relationship of the CH properties and morphology with the corresponding high-speed solar wind profiles.

\begin{acknowledgements}
We would like to thank Drs. R. T. James McAteer and C. Alex Young for helpful discussions.
LDK is a Government of Ireland Scholar and has received funding from the Irish Research Council for Science, Engineering \& Technology (IRCSET) funded by the Irish National Development Plan. Images courtesy of STEREO, SOHO and Hinode. We would like to thank the STEREO/SECCHI consortium for providing open access to their data and technical support. SOHO is a project of international collaboration between ESA and NASA. Hinode is a Japanese mission developed and launched by ISAS/JAXA, with NAOJ as domestic partner and NASA and STFC (UK) as international partners. It is operated by these agencies in co-operation with ESA and NSC (Norway).



\end{acknowledgements}

\bibliographystyle{spr-mp-sola-cnd} \bibliography{sp}

\begin{thebibliography}{49}
\ifx \bisbn   \undefined \def \bisbn  #1{ISBN #1}\fi
\ifx \binits  \undefined \def \binits#1{#1} \fi
\ifx \bauthor  \undefined \def \bauthor#1{#1} \fi
\ifx \batitle  \undefined \def \batitle#1{#1} \fi
\ifx \bjtitle  \undefined \def \bjtitle#1{\textit{#1}}\fi
\ifx \bvolume  \undefined \def \bvolume#1{\textbf{#1}}\fi
\ifx \byear  \undefined \def \byear#1{#1} \fi
\ifx \bissue  \undefined \def \bissue#1{#1} \fi
\ifx \bfpage  \undefined \def \bfpage#1{#1} \fi
\ifx \blpage  \undefined \def \blpage #1{#1} \fi
\ifx \burl  \undefined \def \burl#1{\textsf{#1}} \fi
\ifx \doiurl  \undefined \def \doiurl#1{\textsf{#1}} \fi
\ifx \betal  \undefined \def \betal{\textit{et al.}} \fi
\ifx \binstitute  \undefined \def \binstitute#1{#1} \fi
\ifx \bctitle  \undefined \def \bctitle#1{#1} \fi
\ifx \beditor  \undefined \def \beditor#1{#1} \fi
\ifx \bpublisher  \undefined \def \bpublisher#1{#1} \fi
\ifx \bbtitle  \undefined \def \bbtitle#1{\textit{#1}} \fi
\ifx \bedition  \undefined \def \bedition#1{#1} \fi
\ifx \bseriesno  \undefined \def \bseriesno#1{\textbf{#1}} \fi
\ifx \blocation  \undefined \def \blocation#1{#1} \fi
\ifx \bsertitle  \undefined \def \bsertitle#1{\textit{#1}} \fi
\ifx \bsnm \undefined \def \bsnm#1{#1} \fi
\ifx \bsuffix \undefined \def \bsuffix#1{#1} \fi
\ifx \bparticle \undefined \def \bparticle#1{#1} \fi
\ifx \barticle \undefined \def \barticle#1{#1} \fi
\ifx \botherref \undefined \def \botherref #1{#1} \fi
\ifx \url \undefined \def \url#1{\textsf{#1}} \fi
\ifx \bchapter \undefined \def \bchapter#1{#1} \fi
\ifx \bbook \undefined \def \bbook#1{#1} \fi
\ifx \bcomment \undefined \def \bcomment#1{#1} \fi
\ifx \oauthor \undefined \def \oauthor#1{#1} \fi
\ifx \citeauthoryear \undefined \def \citeauthoryear#1{#1} \fi
\def \endbibitem {}

\bibitem[\protect\citeauthoryear{{Altschuler}, {Trotter}, and
  {Orrall}}{1972}]{Altschuler72}
\begin{barticle}
\bauthor{\bsnm{{Altschuler}}, \binits{M.D.}}, \bauthor{\bsnm{{Trotter}},
  \binits{D.E.}}, \bauthor{\bsnm{{Orrall}}, \binits{F.Q.}}:
\byear{1972},
\bjtitle{\solphys}
\bvolume{26},
\bfpage{354}.
\end{barticle}
\endbibitem

\bibitem[\protect\citeauthoryear{{Antiochos}
  \textit{et~al.}}{2007}]{Antiochos07}
\begin{barticle}
\bauthor{\bsnm{{Antiochos}}, \binits{S.K.}}, \bauthor{\bsnm{{DeVore}},
  \binits{C.R.}}, \bauthor{\bsnm{{Karpen}}, \binits{J.T.}},
  \bauthor{\bsnm{{Miki{\'c}}}, \binits{Z.}}:
\byear{2007},
\bjtitle{\apj}
\bvolume{671},
\bfpage{936}.
\end{barticle}
\endbibitem

\bibitem[\protect\citeauthoryear{{Antonucci}
  \textit{et~al.}}{2004}]{Antonucci04}
\begin{barticle}
\bauthor{\bsnm{{Antonucci}}, \binits{E.}}, \bauthor{\bsnm{{Dodero}},
  \binits{M.A.}}, \bauthor{\bsnm{{Giordano}}, \binits{S.}},
  \bauthor{\bsnm{{Krishnakumar}}, \binits{V.}}, \bauthor{\bsnm{{Noci}},
  \binits{G.}}:
\byear{2004},
\bjtitle{\aap}
\bvolume{416},
\bfpage{749}.
\end{barticle}
\endbibitem

\bibitem[\protect\citeauthoryear{{Belkasim}, {Ghazal}, and
  {Basir}}{2003}]{Belkasim03}
\begin{barticle}
\bauthor{\bsnm{{Belkasim}}, \binits{S.}}, \bauthor{\bsnm{{Ghazal}},
  \binits{A.}}, \bauthor{\bsnm{{Basir}}, \binits{O.A.}}:
\byear{2003},
\bjtitle{Digital Signal Processing}
\bvolume{13},
\bfpage{635}.
\end{barticle}
\endbibitem

\bibitem[\protect\citeauthoryear{{Blush} \textit{et~al.}}{2005}]{Blush05}
\begin{barticle}
\bauthor{\bsnm{{Blush}}, \binits{L.M.}}, \bauthor{\bsnm{{Allegrini}},
  \binits{F.}}, \bauthor{\bsnm{{Bochsler}}, \binits{P.}},
  \bauthor{\bsnm{{Daoudi}}, \binits{H.}}, \bauthor{\bsnm{{Galvin}},
  \binits{A.}}, \bauthor{\bsnm{{Karrer}}, \binits{R.}},
  \bauthor{\bsnm{{Kistler}}, \binits{L.}}, \bauthor{\bsnm{{Klecker}},
  \binits{B.}}, \bauthor{\bsnm{{M{\"o}bius}}, \binits{E.}},
  \bauthor{\bsnm{{Opitz}}, \binits{A.}}, \bauthor{\bsnm{{Popecki}},
  \binits{M.}}, \bauthor{\bsnm{{Thompson}}, \binits{B.}},
  \bauthor{\bsnm{{Wimmer-Schweingruber}}, \binits{R.F.}},
  \bauthor{\bsnm{{Wurz}}, \binits{P.}}:
\byear{2005},
\bjtitle{Adv. Space Res.}
\bvolume{36},
\bfpage{1544}.
\end{barticle}
\endbibitem

\bibitem[\protect\citeauthoryear{{Chapman}}{2007}]{Chapman07}
\begin{botherref}
\oauthor{\bsnm{{Chapman}}, \binits{S.A.}}:
2007,
{The variation of coronal holes with solar cycle}.
PhD thesis,
University of Central Lancashire, Preston, UK.
\end{botherref}
\endbibitem

\bibitem[\protect\citeauthoryear{{Chapman} and {Bromage}}{2002}]{Chapman02}
\begin{botherref}
\oauthor{\bsnm{{Chapman}}, \binits{S.A.}}, \oauthor{\bsnm{{Bromage}},
  \binits{B.J.I.}}:
2002,
In: {Wilson}, A. (ed.)
\textit{From Solar Min to Max: Half a Solar Cycle with SOHO},
\textit{ESA Special Publication}
\textbf{508},
383.
\end{botherref}
\endbibitem

\bibitem[\protect\citeauthoryear{{Chiu} \textit{et~al.}}{1998}]{Chiu98}
\begin{barticle}
\bauthor{\bsnm{{Chiu}}, \binits{M.C.}}, \bauthor{\bsnm{{von-Mehlem}},
  \binits{U.I.}}, \bauthor{\bsnm{{Willey}}, \binits{C.E.}},
  \bauthor{\bsnm{{Betenbaugh}}, \binits{T.M.}}, \bauthor{\bsnm{{Maynard}},
  \binits{J.J.}}, \bauthor{\bsnm{{Krein}}, \binits{J.A.}},
  \bauthor{\bsnm{{Conde}}, \binits{R.F.}}, \bauthor{\bsnm{{Gray}},
  \binits{W.T.}}, \bauthor{\bsnm{{Hunt}}, \binits{J.W.} \bsuffix{Jr.}},
  \bauthor{\bsnm{{Mosher}}, \binits{L.E.}}, \bauthor{\bsnm{{McCullough}},
  \binits{M.G.}}, \bauthor{\bsnm{{Panneton}}, \binits{P.E.}},
  \bauthor{\bsnm{{Staiger}}, \binits{J.P.}}, \bauthor{\bsnm{{Rodberg}},
  \binits{E.H.}}:
\byear{1998},
\bjtitle{Space Scien. Rev.}
\bvolume{86},
\bfpage{257}.
\end{barticle}
\endbibitem

\bibitem[\protect\citeauthoryear{{DeForest}, {Lamy}, and
  {Llebaria}}{2001}]{Deforest01}
\begin{barticle}
\bauthor{\bsnm{{DeForest}}, \binits{C.E.}}, \bauthor{\bsnm{{Lamy}},
  \binits{P.L.}}, \bauthor{\bsnm{{Llebaria}}, \binits{A.}}:
\byear{2001},
\bjtitle{\apj}
\bvolume{560},
\bfpage{490}.
\end{barticle}
\endbibitem

\bibitem[\protect\citeauthoryear{{Delaboudini{\`e}re}
  \textit{et~al.}}{1995}]{Delaboudiniere95}
\begin{barticle}
\bauthor{\bsnm{{Delaboudini{\`e}re}}, \binits{J.P.}},
  \bauthor{\bsnm{{Artzner}}, \binits{G.E.}}, \bauthor{\bsnm{{Brunaud}},
  \binits{J.}}, \bauthor{\bsnm{{Gabriel}}, \binits{A.H.}},
  \bauthor{\bsnm{{Hochedez}}, \binits{J.F.}}, \bauthor{\bsnm{{Millier}},
  \binits{F.}}, \bauthor{\bsnm{{Song}}, \binits{X.Y.}}, \bauthor{\bsnm{{Au}},
  \binits{B.}}, \bauthor{\bsnm{{Dere}}, \binits{K.P.}},
  \bauthor{\bsnm{{Howard}}, \binits{R.A.}}, \bauthor{\bsnm{{Kreplin}},
  \binits{R.}}, \bauthor{\bsnm{{Michels}}, \binits{D.J.}},
  \bauthor{\bsnm{{Moses}}, \binits{J.D.}}, \bauthor{\bsnm{{Defise}},
  \binits{J.M.}}, \bauthor{\bsnm{{Jamar}}, \binits{C.}},
  \bauthor{\bsnm{{Rochus}}, \binits{P.}}, \bauthor{\bsnm{{Chauvineau}},
  \binits{J.P.}}, \bauthor{\bsnm{{Marioge}}, \binits{J.P.}},
  \bauthor{\bsnm{{Catura}}, \binits{R.C.}}, \bauthor{\bsnm{{Lemen}},
  \binits{J.R.}}, \bauthor{\bsnm{{Shing}}, \binits{L.}},
  \bauthor{\bsnm{{Stern}}, \binits{R.A.}}, \bauthor{\bsnm{{Gurman}},
  \binits{J.B.}}, \bauthor{\bsnm{{Neupert}}, \binits{W.M.}},
  \bauthor{\bsnm{{Maucherat}}, \binits{A.}}, \bauthor{\bsnm{{Clette}},
  \binits{F.}}, \bauthor{\bsnm{{Cugnon}}, \binits{P.}}, \bauthor{\bsnm{{van
  Dessel}}, \binits{E.L.}}:
\byear{1995},
\bjtitle{\solphys}
\bvolume{162},
\bfpage{291}.
\end{barticle}
\endbibitem

\bibitem[\protect\citeauthoryear{{Delouille}, {Barra}, and
  {Hochedez}}{2007}]{Barra07}
\begin{botherref}
\oauthor{\bsnm{{Delouille}}, \binits{V.}}, \oauthor{\bsnm{{Barra}},
  \binits{V.}}, \oauthor{\bsnm{{Hochedez}}, \binits{J.}}:
2007,
\textit{AGU Fall Meeting Abstracts}.
SH13A-1107.
\end{botherref}
\endbibitem

\bibitem[\protect\citeauthoryear{{Freeland} and {Handy}}{1998}]{Freeland98}
\begin{barticle}
\bauthor{\bsnm{{Freeland}}, \binits{S.L.}}, \bauthor{\bsnm{{Handy}},
  \binits{B.N.}}:
\byear{1998},
\bjtitle{\solphys}
\bvolume{182},
\bfpage{497}.
\end{barticle}
\endbibitem

\bibitem[\protect\citeauthoryear{{Fujiki} \textit{et~al.}}{2005}]{Fujiki05}
\begin{barticle}
\bauthor{\bsnm{{Fujiki}}, \binits{K.}}, \bauthor{\bsnm{{Hirano}}, \binits{M.}},
  \bauthor{\bsnm{{Kojima}}, \binits{M.}}, \bauthor{\bsnm{{Tokumaru}},
  \binits{M.}}, \bauthor{\bsnm{{Baba}}, \binits{D.}},
  \bauthor{\bsnm{{Yamashita}}, \binits{M.}}, \bauthor{\bsnm{{Hakamada}},
  \binits{K.}}:
\byear{2005},
\bjtitle{Adv. Space Res.}
\bvolume{35},
\bfpage{2185}.
\end{barticle}
\endbibitem

\bibitem[\protect\citeauthoryear{{Gallagher}
  \textit{et~al.}}{1998}]{Gallagher98}
\begin{barticle}
\bauthor{\bsnm{{Gallagher}}, \binits{P.T.}}, \bauthor{\bsnm{{Phillips}},
  \binits{K.J.H.}}, \bauthor{\bsnm{{Harra-Murnion}}, \binits{L.K.}},
  \bauthor{\bsnm{{Keenan}}, \binits{F.P.}}:
\byear{1998},
\bjtitle{\aap}
\bvolume{335},
\bfpage{733}.
\end{barticle}
\endbibitem

\bibitem[\protect\citeauthoryear{{Gallagher}
  \textit{et~al.}}{1999}]{Gallagher99}
\begin{botherref}
\oauthor{\bsnm{{Gallagher}}, \binits{P.T.}}, \oauthor{\bsnm{{Mathioudakis}},
  \binits{M.}}, \oauthor{\bsnm{{Keenan}}, \binits{F.P.}},
  \oauthor{\bsnm{{Phillips}}, \binits{K.J.H.}}, \oauthor{\bsnm{{Tsinganos}},
  \binits{K.}}:
1999,
\textit{\apjl}
\textbf{524}.
L133.
\end{botherref}
\endbibitem

\bibitem[\protect\citeauthoryear{{Golub} \textit{et~al.}}{2007}]{Golub07}
\begin{barticle}
\bauthor{\bsnm{{Golub}}, \binits{L.}}, \bauthor{\bsnm{{Deluca}}, \binits{E.}},
  \bauthor{\bsnm{{Austin}}, \binits{G.}}, \bauthor{\bsnm{{Bookbinder}},
  \binits{J.}}, \bauthor{\bsnm{{Caldwell}}, \binits{D.}},
  \bauthor{\bsnm{{Cheimets}}, \binits{P.}}, \bauthor{\bsnm{{Cirtain}},
  \binits{J.}}, \bauthor{\bsnm{{Cosmo}}, \binits{M.}}, \bauthor{\bsnm{{Reid}},
  \binits{P.}}, \bauthor{\bsnm{{Sette}}, \binits{A.}}, \bauthor{\bsnm{{Weber}},
  \binits{M.}}, \bauthor{\bsnm{{Sakao}}, \binits{T.}}, \bauthor{\bsnm{{Kano}},
  \binits{R.}}, \bauthor{\bsnm{{Shibasaki}}, \binits{K.}},
  \bauthor{\bsnm{{Hara}}, \binits{H.}}, \bauthor{\bsnm{{Tsuneta}},
  \binits{S.}}, \bauthor{\bsnm{{Kumagai}}, \binits{K.}},
  \bauthor{\bsnm{{Tamura}}, \binits{T.}}, \bauthor{\bsnm{{Shimojo}},
  \binits{M.}}, \bauthor{\bsnm{{McCracken}}, \binits{J.}},
  \bauthor{\bsnm{{Carpenter}}, \binits{J.}}, \bauthor{\bsnm{{Haight}},
  \binits{H.}}, \bauthor{\bsnm{{Siler}}, \binits{R.}},
  \bauthor{\bsnm{{Wright}}, \binits{E.}}, \bauthor{\bsnm{{Tucker}},
  \binits{J.}}, \bauthor{\bsnm{{Rutledge}}, \binits{H.}},
  \bauthor{\bsnm{{Barbera}}, \binits{M.}}, \bauthor{\bsnm{{Peres}},
  \binits{G.}}, \bauthor{\bsnm{{Varisco}}, \binits{S.}}:
\byear{2007},
\bjtitle{\solphys}
\bvolume{243},
\bfpage{63}.
\end{barticle}
\endbibitem

\bibitem[\protect\citeauthoryear{{Gonzalez} and {Woods}}{2002}]{Gonzalez02}
\begin{bbook}
\bauthor{\bsnm{{Gonzalez}}, \binits{R.C.}}, \bauthor{\bsnm{{Woods}},
  \binits{R.E.}}:
\byear{2002},
\bbtitle{{Digital Image Processing}},
\bedition{2}nd edn.
\bpublisher{Prentice-Hall},
\blocation{Upper-Saddle River, New Jersey}.
\end{bbook}
\endbibitem

\bibitem[\protect\citeauthoryear{{Guhathakurta} and
  {Fisher}}{1998}]{Guhathakurta98}
\begin{barticle}
\bauthor{\bsnm{{Guhathakurta}}, \binits{M.}}, \bauthor{\bsnm{{Fisher}},
  \binits{R.}}:
\byear{1998},
\bjtitle{\apjl}
\bvolume{499},
\bfpage{215}.
\end{barticle}
\endbibitem

\bibitem[\protect\citeauthoryear{{Harrison}}{1997}]{Harrison97}
\begin{botherref}
\oauthor{\bsnm{{Harrison}}, \binits{R.A.}}:
1997,
In: {Wilson}, A. (ed.)
\textit{Fifth SOHO Workshop: The Corona and Solar Wind Near Minimum Activity}.
SP-404, ESA, Noordwijk, 7.
\end{botherref}
\endbibitem

\bibitem[\protect\citeauthoryear{{Harvey} and {Recely}}{2002}]{Harvey02}
\begin{barticle}
\bauthor{\bsnm{{Harvey}}, \binits{K.L.}}, \bauthor{\bsnm{{Recely}},
  \binits{F.}}:
\byear{2002},
\bjtitle{\solphys}
\bvolume{211},
\bfpage{31}.
\end{barticle}
\endbibitem

\bibitem[\protect\citeauthoryear{{Hassler} \textit{et~al.}}{1999}]{Hassler99}
\begin{barticle}
\bauthor{\bsnm{{Hassler}}, \binits{D.M.}}, \bauthor{\bsnm{{Dammasch}},
  \binits{I.E.}}, \bauthor{\bsnm{{Lemaire}}, \binits{P.}},
  \bauthor{\bsnm{{Brekke}}, \binits{P.}}, \bauthor{\bsnm{{Curdt}},
  \binits{W.}}, \bauthor{\bsnm{{Mason}}, \binits{H.E.}},
  \bauthor{\bsnm{{Vial}}, \binits{J.C.}}, \bauthor{\bsnm{{Wilhelm}},
  \binits{K.}}:
\byear{1999},
\bjtitle{Science}
\bvolume{283},
\bfpage{810}.
\end{barticle}
\endbibitem

\bibitem[\protect\citeauthoryear{{Henney} and {Harvey}}{2005}]{Henney05}
\begin{botherref}
\oauthor{\bsnm{{Henney}}, \binits{C.J.}}, \oauthor{\bsnm{{Harvey}},
  \binits{J.W.}}:
2005,
In: {Sankarasubramanian}, K., {Penn}, M., {Pevtsov}, A. (eds.)
\textit{Large-scale Structures and their Role in Solar Activity}.
CS-346, Astron. Soc. Pac., San Francisco, 261.
\end{botherref}
\endbibitem

\bibitem[\protect\citeauthoryear{{Kano} \textit{et~al.}}{2008}]{Kano08}
\begin{barticle}
\bauthor{\bsnm{{Kano}}, \binits{R.}}, \bauthor{\bsnm{{Sakao}}, \binits{T.}},
  \bauthor{\bsnm{{Hara}}, \binits{H.}}, \bauthor{\bsnm{{Tsuneta}},
  \binits{S.}}, \bauthor{\bsnm{{Matsuzaki}}, \binits{K.}},
  \bauthor{\bsnm{{Kumagai}}, \binits{K.}}, \bauthor{\bsnm{{Shimojo}},
  \binits{M.}}, \bauthor{\bsnm{{Minesugi}}, \binits{K.}},
  \bauthor{\bsnm{{Shibasaki}}, \binits{K.}}, \bauthor{\bsnm{{Deluca}},
  \binits{E.E.}}, \bauthor{\bsnm{{Golub}}, \binits{L.}},
  \bauthor{\bsnm{{Bookbinder}}, \binits{J.}}, \bauthor{\bsnm{{Caldwell}},
  \binits{D.}}, \bauthor{\bsnm{{Cheimets}}, \binits{P.}},
  \bauthor{\bsnm{{Cirtain}}, \binits{J.}}, \bauthor{\bsnm{{Dennis}},
  \binits{E.}}, \bauthor{\bsnm{{Kent}}, \binits{T.}}, \bauthor{\bsnm{{Weber}},
  \binits{M.}}:
\byear{2008},
\bjtitle{\solphys}
\bvolume{249},
\bfpage{263}.
\end{barticle}
\endbibitem

\bibitem[\protect\citeauthoryear{{Kavanagh} and {Denton}}{2007}]{Kavanagh07}
\begin{barticle}
\bauthor{\bsnm{{Kavanagh}}, \binits{A.}}, \bauthor{\bsnm{{Denton}},
  \binits{M.}}:
\byear{2007},
\bjtitle{Astron. and Geophys.}
\bvolume{48}(\bissue{6}),
\bfpage{060000}.
\end{barticle}
\endbibitem

\bibitem[\protect\citeauthoryear{{Krieger}, {Timothy}, and
  {Roelof}}{1973}]{Krieger73}
\begin{barticle}
\bauthor{\bsnm{{Krieger}}, \binits{A.S.}}, \bauthor{\bsnm{{Timothy}},
  \binits{A.F.}}, \bauthor{\bsnm{{Roelof}}, \binits{E.C.}}:
\byear{1973},
\bjtitle{\solphys}
\bvolume{29},
\bfpage{505}.
\end{barticle}
\endbibitem

\bibitem[\protect\citeauthoryear{{Lionello} \textit{et~al.}}{2006}]{Lionello06}
\begin{barticle}
\bauthor{\bsnm{{Lionello}}, \binits{R.}}, \bauthor{\bsnm{{Linker}},
  \binits{J.A.}}, \bauthor{\bsnm{{Miki{\'c}}}, \binits{Z.}},
  \bauthor{\bsnm{{Riley}}, \binits{P.}}:
\byear{2006},
\bjtitle{\apjl}
\bvolume{642},
\bfpage{69}.
\end{barticle}
\endbibitem

\bibitem[\protect\citeauthoryear{{Luhmann} \textit{et~al.}}{2003}]{Luhmann03}
\begin{barticle}
\bauthor{\bsnm{{Luhmann}}, \binits{J.G.}}, \bauthor{\bsnm{{Li}}, \binits{Y.}},
  \bauthor{\bsnm{{Zhao}}, \binits{X.}}, \bauthor{\bsnm{{Yashiro}},
  \binits{S.}}:
\byear{2003},
\bjtitle{\solphys}
\bvolume{213},
\bfpage{367}.
\end{barticle}
\endbibitem

\bibitem[\protect\citeauthoryear{{Malanushenko} and
  {Jones}}{2005}]{Malanushenko05}
\begin{barticle}
\bauthor{\bsnm{{Malanushenko}}, \binits{O.V.}}, \bauthor{\bsnm{{Jones}},
  \binits{H.P.}}:
\byear{2005},
\bjtitle{\solphys}
\bvolume{226},
\bfpage{3}.
\end{barticle}
\endbibitem

\bibitem[\protect\citeauthoryear{{McAteer} \textit{et~al.}}{2005}]{McAteer05}
\begin{barticle}
\bauthor{\bsnm{{McAteer}}, \binits{R.T.J.}}, \bauthor{\bsnm{{Gallagher}},
  \binits{P.T.}}, \bauthor{\bsnm{{Ireland}}, \binits{J.}},
  \bauthor{\bsnm{{Young}}, \binits{C.A.}}:
\byear{2005},
\bjtitle{\solphys}
\bvolume{228},
\bfpage{55}.
\end{barticle}
\endbibitem

\bibitem[\protect\citeauthoryear{{Munro} and {Withbroe}}{1972}]{Munro72}
\begin{barticle}
\bauthor{\bsnm{{Munro}}, \binits{R.H.}}, \bauthor{\bsnm{{Withbroe}},
  \binits{G.L.}}:
\byear{1972},
\bjtitle{\apj}
\bvolume{176},
\bfpage{511}.
\end{barticle}
\endbibitem

\bibitem[\protect\citeauthoryear{{Odstrcil}}{2003}]{Odstrcil03}
\begin{barticle}
\bauthor{\bsnm{{Odstrcil}}, \binits{D.}}:
\byear{2003},
\bjtitle{Adv. Space Res.}
\bvolume{32},
\bfpage{497}.
\end{barticle}
\endbibitem

\bibitem[\protect\citeauthoryear{{Pisanko}}{1997}]{Pisanko97}
\begin{barticle}
\bauthor{\bsnm{{Pisanko}}, \binits{Y.V.}}:
\byear{1997},
\bjtitle{\solphys}
\bvolume{172},
\bfpage{345}.
\end{barticle}
\endbibitem

\bibitem[\protect\citeauthoryear{{Reeves} and {Parkinson}}{1970}]{Reeves70}
\begin{barticle}
\bauthor{\bsnm{{Reeves}}, \binits{E.M.}}, \bauthor{\bsnm{{Parkinson}},
  \binits{W.H.}}:
\byear{1970},
\bjtitle{\apjs}
\bvolume{21},
\bfpage{1}.
\end{barticle}
\endbibitem

\bibitem[\protect\citeauthoryear{{Roussev} \textit{et~al.}}{2003}]{Roussev03}
\begin{barticle}
\bauthor{\bsnm{{Roussev}}, \binits{I.I.}}, \bauthor{\bsnm{{Gombosi}},
  \binits{T.I.}}, \bauthor{\bsnm{{Sokolov}}, \binits{I.V.}},
  \bauthor{\bsnm{{Velli}}, \binits{M.}}, \bauthor{\bsnm{{Manchester}},
  \binits{W.} \bsuffix{IV}}, \bauthor{\bsnm{{DeZeeuw}}, \binits{D.L.}},
  \bauthor{\bsnm{{Liewer}}, \binits{P.}}, \bauthor{\bsnm{{T{\'o}th}},
  \binits{G.}}, \bauthor{\bsnm{{Luhmann}}, \binits{J.}}:
\byear{2003},
\bjtitle{\apjl}
\bvolume{595},
\bfpage{57}.
\end{barticle}
\endbibitem

\bibitem[\protect\citeauthoryear{{Scherrer} \textit{et~al.}}{1995}]{Scherrer95}
\begin{barticle}
\bauthor{\bsnm{{Scherrer}}, \binits{P.H.}}, \bauthor{\bsnm{{Bogart}},
  \binits{R.S.}}, \bauthor{\bsnm{{Bush}}, \binits{R.I.}},
  \bauthor{\bsnm{{Hoeksema}}, \binits{J.T.}}, \bauthor{\bsnm{{Kosovichev}},
  \binits{A.G.}}, \bauthor{\bsnm{{Schou}}, \binits{J.}},
  \bauthor{\bsnm{{Rosenberg}}, \binits{W.}}, \bauthor{\bsnm{{Springer}},
  \binits{L.}}, \bauthor{\bsnm{{Tarbell}}, \binits{T.D.}},
  \bauthor{\bsnm{{Title}}, \binits{A.}}, \bauthor{\bsnm{{Wolfson}},
  \binits{C.J.}}, \bauthor{\bsnm{{Zayer}}, \binits{I.}}, \bauthor{\bsnm{{MDI
  Engineering Team}}}:
\byear{1995},
\bjtitle{\solphys}
\bvolume{162},
\bfpage{129}.
\end{barticle}
\endbibitem

\bibitem[\protect\citeauthoryear{{Scholl} and {Habbal}}{2008}]{Scholl08}
\begin{barticle}
\bauthor{\bsnm{{Scholl}}, \binits{I.F.}}, \bauthor{\bsnm{{Habbal}},
  \binits{S.R.}}:
\byear{2008},
\bjtitle{\solphys}
\bvolume{248},
\bfpage{425}.
\end{barticle}
\endbibitem

\bibitem[\protect\citeauthoryear{{Schrijver} and {DeRosa}}{2003}]{Schrijver03}
\begin{barticle}
\bauthor{\bsnm{{Schrijver}}, \binits{C.J.}}, \bauthor{\bsnm{{DeRosa}},
  \binits{M.L.}}:
\byear{2003},
\bjtitle{\solphys}
\bvolume{212},
\bfpage{165}.
\end{barticle}
\endbibitem

\bibitem[\protect\citeauthoryear{{Schwadron} and {McComas}}{2003}]{Schwadron03}
\begin{barticle}
\bauthor{\bsnm{{Schwadron}}, \binits{N.A.}}, \bauthor{\bsnm{{McComas}},
  \binits{D.J.}}:
\byear{2003},
\bjtitle{\apj}
\bvolume{599},
\bfpage{1395}.
\end{barticle}
\endbibitem

\bibitem[\protect\citeauthoryear{{Sheeley} and {Wang}}{2002}]{Sheeley02}
\begin{barticle}
\bauthor{\bsnm{{Sheeley}}, \binits{N.R.} \bsuffix{Jr.}},
  \bauthor{\bsnm{{Wang}}, \binits{Y.M.}}:
\byear{2002},
\bjtitle{\apj}
\bvolume{579},
\bfpage{874}.
\end{barticle}
\endbibitem

\bibitem[\protect\citeauthoryear{{Temmer}, {Vr{\v s}nak}, and
  {Veronig}}{2007}]{Temmer07}
\begin{barticle}
\bauthor{\bsnm{{Temmer}}, \binits{M.}}, \bauthor{\bsnm{{Vr{\v s}nak}},
  \binits{B.}}, \bauthor{\bsnm{{Veronig}}, \binits{A.M.}}:
\byear{2007},
\bjtitle{\solphys}
\bvolume{241},
\bfpage{371}.
\end{barticle}
\endbibitem

\bibitem[\protect\citeauthoryear{{Timothy}, {Krieger}, and
  {Vaiana}}{1975}]{Timothy75}
\begin{barticle}
\bauthor{\bsnm{{Timothy}}, \binits{A.F.}}, \bauthor{\bsnm{{Krieger}},
  \binits{A.S.}}, \bauthor{\bsnm{{Vaiana}}, \binits{G.S.}}:
\byear{1975},
\bjtitle{\solphys}
\bvolume{42},
\bfpage{135}.
\end{barticle}
\endbibitem

\bibitem[\protect\citeauthoryear{{Toma} and {Arge}}{2005}]{deToma05}
\begin{botherref}
\oauthor{\bsnm{{Toma}}, \binits{G.D.}}, \oauthor{\bsnm{{Arge}}, \binits{C.N.}}:
2005,
In: {Sankarasubramanian}, K., {Penn}, M., {Pevtsov}, A. (eds.)
\textit{Large-scale Structures and their Role in Solar Activity}.
CS-346, Astron. Soc. Pac., San Francisco, 251.
\end{botherref}
\endbibitem

\bibitem[\protect\citeauthoryear{{Vaiana} \textit{et~al.}}{1976}]{Vaiana76}
\begin{barticle}
\bauthor{\bsnm{{Vaiana}}, \binits{G.S.}}, \bauthor{\bsnm{{Zombeck}},
  \binits{M.}}, \bauthor{\bsnm{{Krieger}}, \binits{A.S.}},
  \bauthor{\bsnm{{Timothy}}, \binits{A.F.}}:
\byear{1976},
\bjtitle{\apss}
\bvolume{39},
\bfpage{75}.
\end{barticle}
\endbibitem

\bibitem[\protect\citeauthoryear{{Wang} and {Sheeley}}{1990}]{Wang_Sheeley90}
\begin{barticle}
\bauthor{\bsnm{{Wang}}, \binits{Y.M.}}, \bauthor{\bsnm{{Sheeley}},
  \binits{N.R.} \bsuffix{Jr.}}:
\byear{1990},
\bjtitle{\apj}
\bvolume{355},
\bfpage{726}.
\end{barticle}
\endbibitem

\bibitem[\protect\citeauthoryear{{Wang}, {Hawley}, and
  {Sheeley}}{1996}]{Wang96}
\begin{barticle}
\bauthor{\bsnm{{Wang}}, \binits{Y.M.}}, \bauthor{\bsnm{{Hawley}},
  \binits{S.H.}}, \bauthor{\bsnm{{Sheeley}}, \binits{N.R.} \bsuffix{Jr.}}:
\byear{1996},
\bjtitle{Science}
\bvolume{271},
\bfpage{464}.
\end{barticle}
\endbibitem

\bibitem[\protect\citeauthoryear{{Wiegelmann} and
  {Solanki}}{2004}]{Wiegelmann04}
\begin{botherref}
\oauthor{\bsnm{{Wiegelmann}}, \binits{T.}}, \oauthor{\bsnm{{Solanki}},
  \binits{S.K.}}:
2004,
In: {Walsh}, R.W., {Ireland}, J., {Danesy}, D., {Fleck}, B. (eds.)
\textit{SOHO 15 Coronal Heating}.
SP-575, ESA, Noordwijk, 35.
\end{botherref}
\endbibitem

\bibitem[\protect\citeauthoryear{{Wiegelmann}, {Xia}, and
  {Marsch}}{2005}]{Wiegelmann05}
\begin{barticle}
\bauthor{\bsnm{{Wiegelmann}}, \binits{T.}}, \bauthor{\bsnm{{Xia}},
  \binits{L.D.}}, \bauthor{\bsnm{{Marsch}}, \binits{E.}}:
\byear{2005},
\bjtitle{\aap}
\bvolume{432},
\bfpage{1}.
\end{barticle}
\endbibitem

\bibitem[\protect\citeauthoryear{{Wilhelm}}{2006}]{Wilhelm06}
\begin{barticle}
\bauthor{\bsnm{{Wilhelm}}, \binits{K.}}:
\byear{2006},
\bjtitle{\aap}
\bvolume{455},
\bfpage{697}.
\end{barticle}
\endbibitem

\bibitem[\protect\citeauthoryear{{Wuelser} \textit{et~al.}}{2004}]{Wuelser04}
\begin{botherref}
\oauthor{\bsnm{{Wuelser}}, \binits{J.P.}}, \oauthor{\bsnm{{Lemen}},
  \binits{J.R.}}, \oauthor{\bsnm{{Tarbell}}, \binits{T.D.}},
  \oauthor{\bsnm{{Wolfson}}, \binits{C.J.}}, \oauthor{\bsnm{{Cannon}},
  \binits{J.C.}}, \oauthor{\bsnm{{Carpenter}}, \binits{B.A.}},
  \oauthor{\bsnm{{Duncan}}, \binits{D.W.}}, \oauthor{\bsnm{{Gradwohl}},
  \binits{G.S.}}, \oauthor{\bsnm{{Meyer}}, \binits{S.B.}},
  \oauthor{\bsnm{{Moore}}, \binits{A.S.}}, \oauthor{\bsnm{{Navarro}},
  \binits{R.L.}}, \oauthor{\bsnm{{Pearson}}, \binits{J.D.}},
  \oauthor{\bsnm{{Rossi}}, \binits{G.R.}}, \oauthor{\bsnm{{Springer}},
  \binits{L.A.}}, \oauthor{\bsnm{{Howard}}, \binits{R.A.}},
  \oauthor{\bsnm{{Moses}}, \binits{J.D.}}, \oauthor{\bsnm{{Newmark}},
  \binits{J.S.}}, \oauthor{\bsnm{{Delaboudiniere}}, \binits{J.P.}},
  \oauthor{\bsnm{{Artzner}}, \binits{G.E.}}, \oauthor{\bsnm{{Auchere}},
  \binits{F.}}, \oauthor{\bsnm{{Bougnet}}, \binits{M.}},
  \oauthor{\bsnm{{Bouyries}}, \binits{P.}}, \oauthor{\bsnm{{Bridou}},
  \binits{F.}}, \oauthor{\bsnm{{Clotaire}}, \binits{J.Y.}},
  \oauthor{\bsnm{{Colas}}, \binits{G.}}, \oauthor{\bsnm{{Delmotte}},
  \binits{F.}}, \oauthor{\bsnm{{Jerome}}, \binits{A.}},
  \oauthor{\bsnm{{Lamare}}, \binits{M.}}, \oauthor{\bsnm{{Mercier}},
  \binits{R.}}, \oauthor{\bsnm{{Mullot}}, \binits{M.}},
  \oauthor{\bsnm{{Ravet}}, \binits{M.F.}}, \oauthor{\bsnm{{Song}},
  \binits{X.}}, \oauthor{\bsnm{{Bothmer}}, \binits{V.}},
  \oauthor{\bsnm{{Deutsch}}, \binits{W.}}:
2004,
In: {Fineschi}, S., {Gummin}, M.A. (eds.)
\textit{Telescopes and Instrumentation for Solar Astrophysics}.
CS-5171, SPIE, 111.
\end{botherref}
\endbibitem

\end{thebibliography}

\end{article} \end{document}